\documentclass{aa}  

\usepackage[colorlinks = true,
            linkcolor = blue,
            urlcolor  = blue,
            citecolor = blue,
            anchorcolor = blue]{hyperref}

\usepackage{amsmath}
\usepackage{color}
\usepackage{graphicx}
\usepackage{txfonts}

\newcommand{\insitu}{\textit{in situ}}

\usepackage{amsmath}
\usepackage{array}
\usepackage{tabulary}
\usepackage{rotating}
\usepackage{lineno}

\graphicspath{{./}{figures/}}

\begin{document}

	 \title{Total Electron Temperature Derived from Quasi-Thermal Noise Spectroscopy In the Pristine Solar Wind: Parker Solar Probe Observations}
	 \titlerunning{Radial Evolution of ${T}_{e}$ Derived from QTN Technique}

	 \author{Mingzhe Liu\inst{1}
					\and Karine Issautier\inst{1}
					\and Michel Moncuquet\inst{1}
					\and Nicole Meyer-Vernet\inst{1}
					\and Milan Maksimovic\inst{1}
					\and Jia Huang\inst{2}
					\and Mihailo M. Martinovic\inst{3,1}
					\and Léa Griton\inst{1}
					\and Nicolina Chrysaphi\inst{1}
					\and Vamsee Krishna Jagarlamudi\inst{4}
					\and Stuart D. Bale\inst{5,6}
					\and Marc Pulupa\inst{5}
					\and Justin C. Kasper\inst{2,7}
					\and M. L. Stevens\inst{7}
					}

	 \institute{LESIA, Observatoire de Paris, Université PSL, CNRS, 
Sorbonne Université, Université de Paris, 
5 place Jules Janssen, 92195 Meudon, France\\
							\email{mingzhe.liu@obspm.fr}
				 \and Climate and Space Sciences and Engineering, University of Michigan, Ann Arbor, MI 48109, USA
				 \and Lunar and Planetary Laboratory, University of Arizona, Tucson, AZ 85721, USA
				 \and Johns Hopkins University Applied Physics Laboratory, Laurel, MD, USA
				 \and Space Sciences Laboratory, University of California, Berkeley, CA 94720-7450, USA
  				 \and Physics Department, University of California, Berkeley, CA 94720-7300, USA
				 \and Smithsonian Astrophysical Observatory, Cambridge, MA 02138 USA	   
				}

	 \date{Received 2022 November 12; Accepted 2023 March 15}

	\abstract
	 {} 
	 {We apply the Quasi-thermal noise (QTN) method on Parker Solar Probe (PSP) observations to derive the total electron temperature ($T_e$) and present a combination of 12-day period of observations around each perihelion from Encounter One (E01) to Ten (E10) (with E08 not included) with the heliocentric distance varying from about 13 to 60 solar radii ($R_{\sun}$).} 
	 {The QTN technique is a reliable tool to yield accurate measurements of the electron parameters in the solar wind. We obtain $T_e$ from the linear fit of the high-frequency part of the QTN spectra acquired by the RFS/FIELDS instrument. Then, we provide the mean radial electron temperature profile, and examine the electron temperature gradients for different solar wind populations (i.e. classified by the proton bulk speed ($V_p$), and the solar wind mass flux).}
	 {We find that the total electron temperature decreases with the distance as $\sim$$R^{-0.66}$, which is much slower than adiabatic. The extrapolated $T_e$ based on PSP observations is consistent with the exospheric solar wind model prediction at $\sim$10 $R_{\sun}$, Helios observations at $\sim$0.3 AU and Wind observations at 1 AU. Also, $T_e$, extrapolated back to 10 $R_{\sun}$, is almost the same as the strahl electron temperature $T_s$ (measured by SPAN-E) which is considered to be closely related to or even almost equal to the coronal electron temperature. Furthermore, the radial $T_e$ profiles in the slower solar wind (or flux tube with larger mass flux) are steeper than those in the faster solar wind (or flux tube with smaller mass flux). More pronounced anticorrelated $V_p$--$T_e$ is observed when the solar wind is slower and closer to the Sun.}
	 {}


	 \maketitle
%

\section{Introduction} \label{introduction}


Heat transport in the solar corona and wind, which is not completely understood, plays a key role in coronal heating and wind acceleration. Due to the large mass difference between ions and electrons, electrons mainly transport energy whereas ions transport momentum. Therefore, electrons are expected to play a key role in the thermally driven solar wind expansion. Furthermore, the accurately measured electron temperature radial profile is not only of prime interest to understand the energy transport in the solar wind but also an important ingredient to constrain the thermally driven solar wind models \citep[e.g.,][]{1998meyervernet,1999Issautierb,2001Issautierb,1997Maksimovic,2004Zouganelis}. For simplicity, the electron temperature is generally assumed to be fitted with a power law of the distance to the Sun, assuming no large--scale temporal variations: $T_e = T_0\times(R/{R_{\sun}})^{\beta}$. $\beta$ is observed to range between 0 (isothermal) and $-4/3$ (adiabatic), which indicates that electrons cool off with radial profiles spanning from nearly isothermal to almost adiabatic \citep[e.g.,][]{1989Marsch,1990Pilipp,1998Issautier,2011LeChat,2000Maksimovic,2015JGRA,2020Moncuquet}. The large scatter in the measurements of $\beta$ is not surprising and may be due to several reasons: i) it is difficult to separate genuine variations along stream flux tubes from those across them; ii) transient structures such as coronal mass ejections, co-rotating interaction regions and interplanetary shocks can cause nongeneric effects; iii) the observations from different spacecraft have been carried out in different latitudinal and radial ranges and/or in different phases of the solar activity; iv) classification of data based on the solar wind speed, Coulomb collisions and plasma beta has not always been done. In contrast, the exospheric solar wind models give another theoretical radial profile of the total electron temperature with the expression 
$T_e = T_0 +T_1\times(R/{R_{\sun}})^{-4/3}$ for $(R/{R_{\sun}})^2 \gg 1$ \citep[e.g.,][]{1998meyervernet,2003meyervernet,2001Issautierb}, which yields a profile that flattens at large distances, in agreement with Helios measurements (between 0.3 and 1 AU) \citep{1989Marsch,1990Pilipp}. Since this model has the same number of free parameters as the power--law model, it is difficult to distinguish both models from observations in a small radial range. \cite{2011LeChat} has verified this fact with the Ulysses observations of high-speed solar wind during its first pole-to-pole latitude scan (from 1.5 to 2.3 AU).

Observations from Parker Solar Probe \citep[PSP;][]{2016Fox} indicate that there is an anticorrelation between the proton bulk speed $V_p$ and the electron temperature $T_e$ close to the Sun \citep[e.g.,][]{Maksimovic2020, 2020Halekasa, 2022Halekas}, whereas the correlation between the proton bulk speed $V_p$ and the proton temperature $T_p$ persists throughout the heliosphere \citep[see][and references therein]{1986Lopez,1995Totten,2006Matthaeus,2009Demoulin}. Specifically, \cite{Maksimovic2020} found that the anticorrelation between $V_p$ and $T_e$ observed below 0.3 AU disappears as the wind expands, evolves and mixes with different electron temperature gradients for different wind speeds. The exospheric solar wind model \citep[e.g.,][]{1997Maksimovic, 2001Maksimovic} showed that the fast wind from the polar coronal hole regions (low-temperature regions) might be produced by the non-thermal electron distributions in the corona, which might explain the anticorrelated $(V_{p}, T_{e})$ close to the Sun. Furthermore, the exospheric model predicted that the temperature profile is flatter in the fast wind as previously observed \citep{1998meyervernet}. However, exospheric models use simplified hypotheses and challenging questions remain about the heating and cooling mechanisms for electrons. The PSP observations close to the Sun therefore give us an opportunity to investigate the solar wind electron thermal dynamics in the inner heliosphere.

The Quasi-thermal noise (QTN) technique yields accurate electron density and temperature measurements in the solar wind. It has been used in a number of space missions \citep[e.g.,][]{1979meyervernet,1986meyervernet,1993MeyerVernet,2017meyervernet,1999Issautier,2001Issautier,2001Issautierc,2005Issautier,2008Issautier,1995Maksimovic,2005Maksimovic_AdSpR,1995Moncuquet,1997Moncuquet,2005Moncuquet,2006Moncuquet,2020Martinovic,2011LeChat,2001salem,1994Lund,2013Schippers}. Recent investigations \citep[see][]{2020Moncuquet, Maksimovic2020,2022Martinovic} have already applied this technique on PSP based on electric voltage spectra acquired by the Radio Frequency Spectrometer (RFS/FIELDS) \citep{2017Pulupa}. Besides, SWEAP/PSP consists of the Solar Probe Cup (SPC) and the Solar Probe Analyzers (SPAN) \citep{Kasper2016, Case2020, Whittlesey_2020, 2022Livi}. SPC is a fast Faraday cup designed to measure the one dimensional velocity distribution function (VDF) of ions. SPAN is a combination of three electrostatic analyzers operated to measure the three dimensional ion and electron VDFs. Usually, traditional particle analyzers are affected by spacecraft photoelectrons and charging effects. Since the QTN electron density is deduced from a spectral peak, this measurement is independent of gain calibrations. Due to its reliability and accuracy, the electron number density derived from the QTN spectroscopy is called the gold standard density and serves routinely to calibrate other instruments \citep[e.g.,][]{1995Maksimovic,2001Issautierc,2001salem}. Until now, on PSP, electron number density provided by the QTN technique has been playing an important role as a calibration standard for the scientific analysis \citep[e.g.,][]{2021Kasper,2021Zhao,2021Liu,2021ApJLiu}.

We derived the total electron temperature from the QTN spectroscopy in the so-called pristine or nascent solar wind observed by PSP. Specifically, a combination of 12-day period of observations around each perihelion from Encounter One (E01) to Ten (E10) of PSP are presented with the heliocentric distance varying from about 13 to 60 $R_\odot{}$. Currently, observations from E08 are not considered due to the unusual biasing setting for the electric antenna at that time. In Section \ref{overview}, we describe a simple but practical and effective way to deduce the total electron temperature with the high-frequency part of the quasi-thermal noise spectra provided by the dipole electric antenna onboard PSP. The corresponding results are compared to those from \cite{Maksimovic2020} ($T_e$ from a different QTN technique), and \cite{2020Moncuquet} ($T_c$ from a simplified QTN technique) for a preliminary cross-checking. In Section \ref{3}, we first provide the mean radial electron temperature profile, and then investigate the electron temperature gradients for different solar wind populations classified by the proton bulk speed and the solar wind mass flux, respectively. Also, we examine how the anticorrelation between $V_p$ and $T_e$ are affected by radial evolution. In Section \ref{summary_and_discussion}, the results and their implications for the electron thermal dynamics are summarized and discussed. 

\section{Data analysis} \label{overview}

The QTN spectroscopy technique provides in situ macroscopic plasma properties by analyzing the power spectrum of the electric field voltage induced on an electric antenna by the plasma particle quasi-thermal motions. The QTN spectra are determined by both the ambient plasma properties and the antenna configuration because of the strong coupling between the plasma particles and the electric field. For an ideal electric antenna configuration, the longer and thinner the electric antenna is set, the better the QTN technique performs. Specifically, the length of the electric antenna ($L$) should exceed the local Debye length $L_D$ to ensure accurate temperature measurements. Fat antennas (with thick radius $a$) collect or emit more electrons so that the corresponding shot noise may exceed the quasi-thermal noise. Therefore, the electric antenna should be both long enough and thin enough ($a$ $\textless$ $L_D$ $\textless$ $L$) so that the QTN technique can work well. When the antenna is configured in a proper way \citep[see][]{2017meyervernet,2020meyervernet}, the QTN spectra are completely determined by the particle velocity distributions of the ambient plasma. 

The QTN spectrum around the electron plasma frequency ($f_p$) contains a wealth of infomation about the solar wind, whose basic shape can be explained based on simple plasma physics \citep{1989meyervernet}. The quasi-thermal motion of the ambient plasma electrons passing by the antenna induces electric voltage pulses. At time scales exceeding $1$/$(2 \pi f_p)$ (corresponding to frequencies $f$ $\textless$ $f_p$), the electrons are Debye shielded so that each ambient thermal electron passing closer than $L_D$ produces on the antenna an electric voltage pulse with a duration (roughly equal to $1$/$(2 \pi f_p)$) shorter than the inverse frequency of observation. Thus, the Fourier transform of such a pulse is a constant for $f$ $\textless$ $f_p$, producing a plateau whose amplitude is determined by the bulk of the thermal electrons. In contrast, at higher frequencies ($f$ $\textgreater$ $f_p$), the electron quasi-thermal motion excites Langmuir waves, thereby producing a spectral peak near $f_p$ as well as a power spectrum proportional to the total electron pressure at high frequencies \citep{1989meyervernet,2017meyervernet}. Furthermore, the height of the peak near $f_p$ depends on the mean energy of suprathermal electrons, whereas the peak width depends on suprathermal electron concentration \citep{1991Chateau,2017meyervernet}.

For the first several encounters, PSP is still not close enough to the Sun and therefore $L_D$ generally exceeds the antenna length ($L\simeq2$ m). Nevertheless, the plasma peak emerged because of the suprathermal electrons \citep{2022meyervernet}. Therefore, \cite{2020Moncuquet} gave the first results of QTN measurements on PSP based on a simplified QTN technique. The preliminary results include the electron number density $n_e$, the core electron temperature $T_c$ and an estimation of the suprathermal electron temperature $T_h$ (contribution of both the halo and strahl electron thermal pressure). Based on the derived $n_e$ from \cite{2020Moncuquet}, \cite{Maksimovic2020} yields the total electron temperature $T_e$ during the first encounter of PSP by fitting the high-frequency part of the QTN spectra recorded by RFS/FIELDS. In this paper, we apply another simple, fast but effective method on PSP observations to derive $T_e$. In the next subsection, we provide details of the method that enables us to derive the total electron temperature from the high-frequency part of the electric field voltage spectra measured by RFS/FIELDS. Finally, we present the preliminary cross-checking between the total electron temperature derived in this work and those obtained via different QTN techniques.

\subsection{Determination of $T_e$ from QTN Spectroscopy} \label{particle_distribution}

In practice, the measured electric field voltage power spectrum at the receiver ports is expressed as

\begin{equation}\label{e1}
\begin{aligned}
{V^2_{R}} & = {\Gamma^2_R}({V^2_{electron}}+{V^2_{proton}}+{V^2_{shot}})+{V^2_{noise}}+{V^2_{galaxy}}
\end{aligned}
\end{equation}

where $V^2_{electron}$, $V^2_{proton}$, $V^2_{shot}$, $V^2_{noise}$, and $V^2_{galaxy}$ represent the electron QTN, the doppler-shifted proton thermal noise, the shot noise, the instrument noise, and the galactic radio background noise, respectively. In Equation \ref{e1}, $\Gamma^2_R$ is the gain factor of the receiver, which is expressed as

\begin{equation}\label{e2}
\begin{aligned}
\Gamma^2_R & \simeq \frac{C^2_A}{{(C_A+C_B)}^2}
\end{aligned}
\end{equation}

where $C_{A}$ and $C_{B}$ are the dipole antenna capacitance and the (dipole) stray capacitance, respectively. Since $V^2_{R}$ is the power spectrum at the receiver ports, $\Gamma^2_R$ is in factor of the first three terms. Note that $\Gamma^2_R$ has already been included in the expression of $V^2_{galaxy}$ (see below). For the frequencies satisfying $fL/(f_pL_D)\gg1$, the electron QTN can be approximated as $V^2_{electron} \simeq \frac{f^2_pk_BT_e}{\pi\epsilon_0L^{'}f^3}$ \citep{1989meyervernet}, where $f_p$ is the local electron plasma frequency, $L^{'}$ equals to the physical length ($L$) of one boom (or arm) of the dipole antenna when it is long enough (i.e. $L \gg L_D$), $k_B$ is the Boltzmann constant, and $\epsilon_0$ is the permittivity of free space. PSP/FIELDS antennas are separated by the heat shield and the physical separation is $\sim$3 meters for both $\lvert$$V1$$-$$V2$$\rvert$ and $\lvert$$V3$$-$$V4$$\rvert$ dipole antennas. Since the antenna physical length ($L\simeq2$ m) is not long enough, the gap should be considered for $L^{'}$ with $L^{'}=3.5$ m. The high-frequency electron QTN (above $f_p$) is proportional to the electron kinetic temperature whatever the shape of the velocity distribution is like. For the frequency ranges considered, $C_{A}\simeq\pi\epsilon_0L/[ln(L/a)-1]$ \citep{2017meyervernet} and $C_{B}\simeq18$ pF \citep{2020Moncuquet}, where $L\simeq2$ m is the electric antenna physical length and $a\simeq1.5$ mm is the wire radius. Note that, when performing the fitting using the whole QTN spectra, the derived electron temperatures depend on the choice of the velocity distribution function for the electrons \citep{1989meyervernet}. This is similar to the analysis to fit the velocity distribution functions observed by the particle analyzer. However, in the present work, the derived total electron temperature is not model dependent. This is because, when deriving the expression of $V^2_{electron} \simeq \frac{f^2_pk_BT_e}{\pi\epsilon_0L^{'}f^3}$, $T_e$ is defined directly from the second moment of the electron velocity distribution functions, and no models are assumed \citep{1989meyervernet,1991Chateau,2017meyervernet}.

When $fL/(f_pL_D)\gg1$, the the doppler-shifted proton thermal noise and the shot noise are negligible compared to the electron QTN \citep{2017meyervernet}. Note that the periodic antenna biasing performed for measuring the DC electric fields, which affect the shot noise, do not perturb our results, contrary to the perturbations these biasing bring to the QTN at smaller frequencies. In contrast, the contributions of the galactic radio background noise \citep{1978Novaco,1979Cane,2011Zaslavsky} and the instrument noise become important and need to be substracted to obtain the effective electron QTN spectrum at high frequency. The galaxy noise is almost constant in time and nearly isotropic in angular distribution with the modulation as a function of the observed solid angle being less than 20$\%$ in the considered frequency range \citep{2001Manning}. Therefore, it was frequently used to calibrate the antenna onboard previous spacecraft missions \citep[e.g.,][]{2011Zaslavsky,Maksimovic2020}. Specifically, the calibration is performed by relating the measured radio background radiation of the galaxy to the modelled flux of the source. The use of the empirical isotropic galaxy noise model from \cite{1978Novaco} was justified by displaying a good agreement between the data and the model. Due to the high sensitivity of RFS/FIELDS/PSP \citep{2017Pulupa}, the galaxy noise lies within the RFS bandwidth and can be accurately measured. As a result, following the method outlined in \cite{2011Zaslavsky}, \cite{Maksimovic2020} used an RFS spectrum measured when PSP was close to 1 AU to derive an accurate absolute value of the reduced effective length of $\lvert$$V1$$-$$V2$$\rvert$ dipole antenna. Below, the galaxy noise measured by RFS/FIELDS/PSP is modelled based on the newly derived reduced effective length of $\lvert$$V1$$-$$V2$$\rvert$ dipole antenna. The pre-deployment internal noise of RFS/FIELDS (after launch) in the considered frequency range was estimated to be $V^2_{noise}$$\sim$ 2.2$\times$10$^{-17}$ V$^2$Hz$^{-1}$ \citep{2020Pulupa,Maksimovic2020}. The background radio galactic noise is modelled following the procedures of \cite{2011Zaslavsky} and \cite{Maksimovic2020}. The specific steps are summarized below. 

The background radio galactic noise is modelled according to Equation (11) from \cite{2011Zaslavsky}, $V^2_{galaxy}=\frac{4\pi}{3}Z_0\Gamma^2_RL^2_{eff}B_{model}$, where $Z_0=\sqrt{\mu_0/\epsilon_0}\simeq120\pi$ is the impedance of vacuum, $\Gamma_RL_{eff}=1.17$ is the reduced effective length \citep[see][]{Maksimovic2020}, and $B_{model}$ is the empirical model for the isotropic sky background brightness \citep{1978Novaco}, expressed as
\begin{equation}\label{e3}
\begin{aligned}
B_{model} & = B_0f^{-0.76}_{MHz}e^{-\tau}
\end{aligned}
\end{equation}
where $B_{0}=1.38 \times 10^{-19} W/m^2/Hz/sr$, $f_{MHz}$ is the frequency expressed in MHz, and $\tau=3.28f^{-0.64}_{MHz}$.

\begin{figure}
	\centering
        \includegraphics[width=0.5\textwidth]{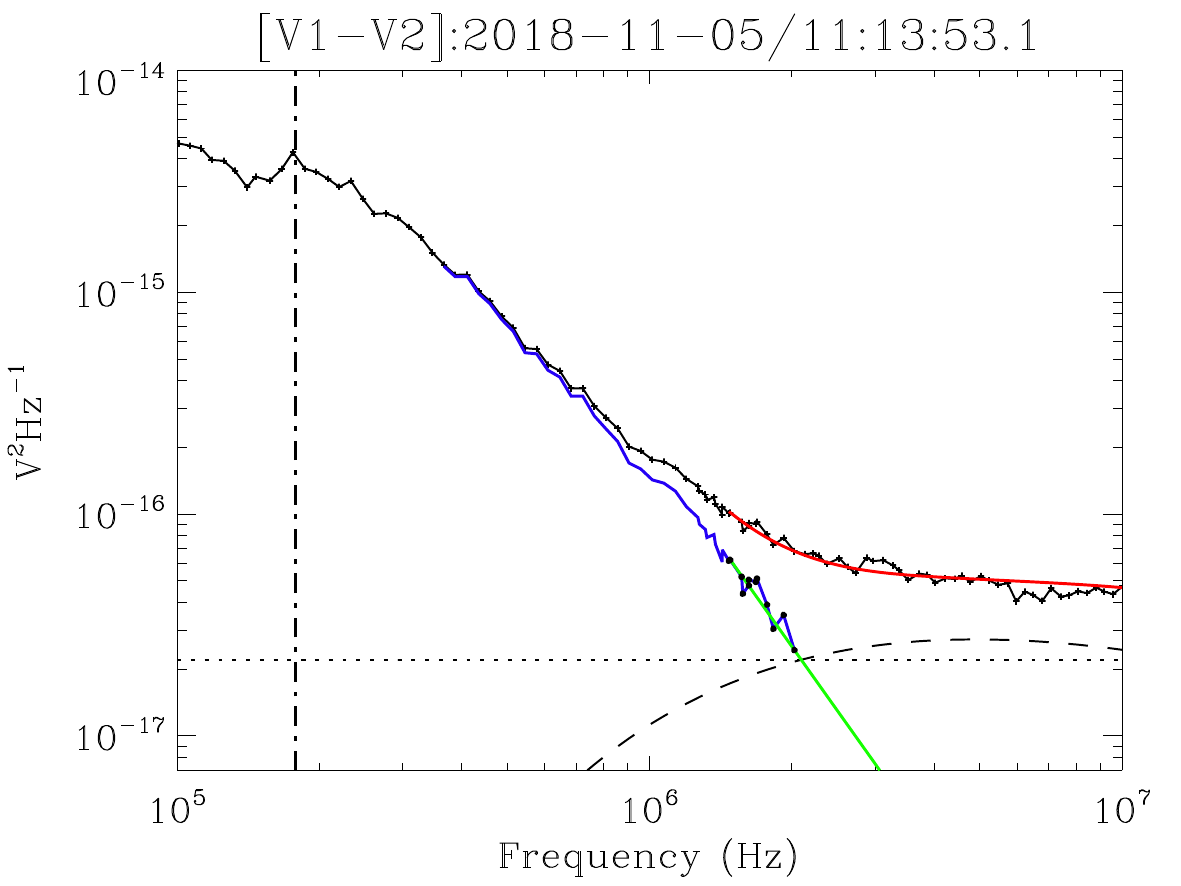}
	\caption{Example of a voltage power spectrum (between 100 kHz and 10 MHz) recorded by the RFS receiver using the $\lvert$$V1$$-$$V2$$\rvert$ dipole electric antennas from FIELDS instrumentation (full black curve connected by crosses). The dot--dashed line gives the position of the local plasma peak \citep{2020Moncuquet}. The dotted horizontal line represents the pre-deployment RFS instrument noise (after launch) of $\sim$2.2 $\times$ 10$^{-17} $V$^2$Hz$^{-1}$. The black dashed line shows the modelled radio galaxy noise. The blue and green lines represent the effective QTN signal and linear fit ($f^{-3}$ variation that the QTN spectrum should follow when $f \gg f_{p}L_{D}/L$), respectively. The black dots on the blue line are used to derive the linear fit. The red line is the sum of the fitted QTN signal, the modelled radio galaxy noise and the instrument noise. The details are described in the text.} 
\label{f1}
\end{figure}

Figure~\ref{f1} presents a typical example of electric field voltage power spectrum plot ranging between 100 kHz and 10 MHz measured by the FIELDS $\lvert$$V1$$-$$V2$$\rvert$ dipole antenna connected to the RFS receiver. We obtain the observations (crosses connected by the black curve) by merging the spectra measured by LFR/RFS and that by HFR/RFS. The dotted horizontal line represents pre-deployment internal noise of RFS/FIELDS as mentioned above. The black dashed line is the radio galaxy background noise calculated as described above. Both the intrument noise and radio galaxy noise are deducted from the observed electric field voltage spectrum so that the so-called pure QTN spectrum ${\Gamma^2_R}{V^2_{electron}}\simeq{V^2_{R}}-{V^2_{noise}}-{V^2_{galaxy}}$ (blue curve line) is derived following the similar requirement set by \cite{Maksimovic2020}. Specifically, the data points are selected as: (1) the lower-frequency limit is set as $fL/(f_pL_D)\ge2$ so that both the proton thermal noise and the shot noise can be neglected; (2) the derived so-called pure QTN spectrum should be larger than both the instrument noise and radio galaxy noise, which is used to set the higher-frequency limit. Then, we further select the dataset for the linear fitting to derive $T_e$ following $fL/(f_pL_D)\ge8$, which is a much more strict requirement. The green line represents the linear fitted results and there is only one free parameter which is the total electron temperature. The electron plasma frequency used for each fitted spectrum is derived from the plasma peak tracking technique \citep[see][]{2020Moncuquet}. In Figure~\ref{f1}, the vertical black dashed-dotted line, which represents the location of the local electron plasma frequency, is plotted for reference. Specifically, we perform the numerical process by fitting the theoretical voltage spectral density $ log(V^2_{R})$ to each measured spectrum via minimizing the $\chi^2$ merit function with the implementation of a nonlinear least-squares Levenberg-Marquardt technique \citep{2009Markwardt}. $\chi^2$ is defined as $\chi^2=\sum_{i=1}^N \frac{(O_i-E_i)^2}{O_i}$, where $O_i$ is the value of the measured spectrum, and $E_i$ is the corresponding expected value (theoretical one). All the electric field voltage spectra measured by the RFS/FIELDS are fitted following the same procedure mentioned above. The spectra fitted in this work usually comprise a number of frequency points ranging between $\sim$5 and $\sim$15. We further quantify the quality of the fit with the overall standard deviation ($\sigma_{fit}$) of the numerical fitted values to the corresponding measurements. In general, $\sigma_{fit}$ $\textless$ 2.5$\%$ indicates the goodness of the fittings. The physical uncertainty of $T_e$ is estimated from the uncertainty of the plasma frequency, the uncertainty of the so-called pure QTN spectrum and the uncertainty of the numerical process. The uncertainty of the plasma frequency is about 4$\%$ ($\sim$8$\%$ for $f^2_p$) \citep{2020Moncuquet}, which is the standard frequency resolution of the RFS/FIELDS. The uncertainty of the so-called pure QTN spectrum comes from the variations of the instrument noise and the empirical isotropic galaxy noise model, which is in total less than 20$\%$ \citep[see][]{2001Manning,2011Zaslavsky,2017Pulupa,Maksimovic2020}. The uncertainty for the sum of the instrument noise and the empirical isotropic galaxy noise model mainly affect the pure QTN spectrum at the highest selected frequency, whereas it is negligible at the lowest selected frequency. For simplicity, the mean uncertainty of the so-called pure QTN spectrum is estimated to be about 10$\%$. Therefore, the physical uncertainty of the derived $T_e$ is at most 20$\%$, which is almost the same as that of $T_c$ \citep[see][]{2020Moncuquet,2020Liu,2021Liu}. This estimated physical uncertainty for $T_e$ is consistent with the statistical uncertainty for $T_e$ shown in Figure~\ref{f4}.

Note that the high-frequency part of the QTN spectrum can be strongly perturbed by the electromagnetic emissions (e.g., Type II and/or III radio emissions) and sometimes cannot be used for deriving $T_e$. Especially, such electromagnetic emissions were frequently detected during E02 \citep{2020Pulupa} and should be carefully removed. In this work, when the QTN technique cannot be implemented in the presence of electromagnetic emissions, no $T_e$ value is derived. The electric field voltage power spectrum below $f_p$ in general remains unperturbed and both $n_e$ and $T_c$ can still be obtained \citep{2020Moncuquet}. As a byproduct, we managed to derive a database of spectra affected by bursty Langmuir waves and/or electromagnetic emissions.

\subsection{Preliminary Cross Checking} \label{cross-checking}

\begin{figure}
	\centering
        \includegraphics[width=0.5\textwidth]{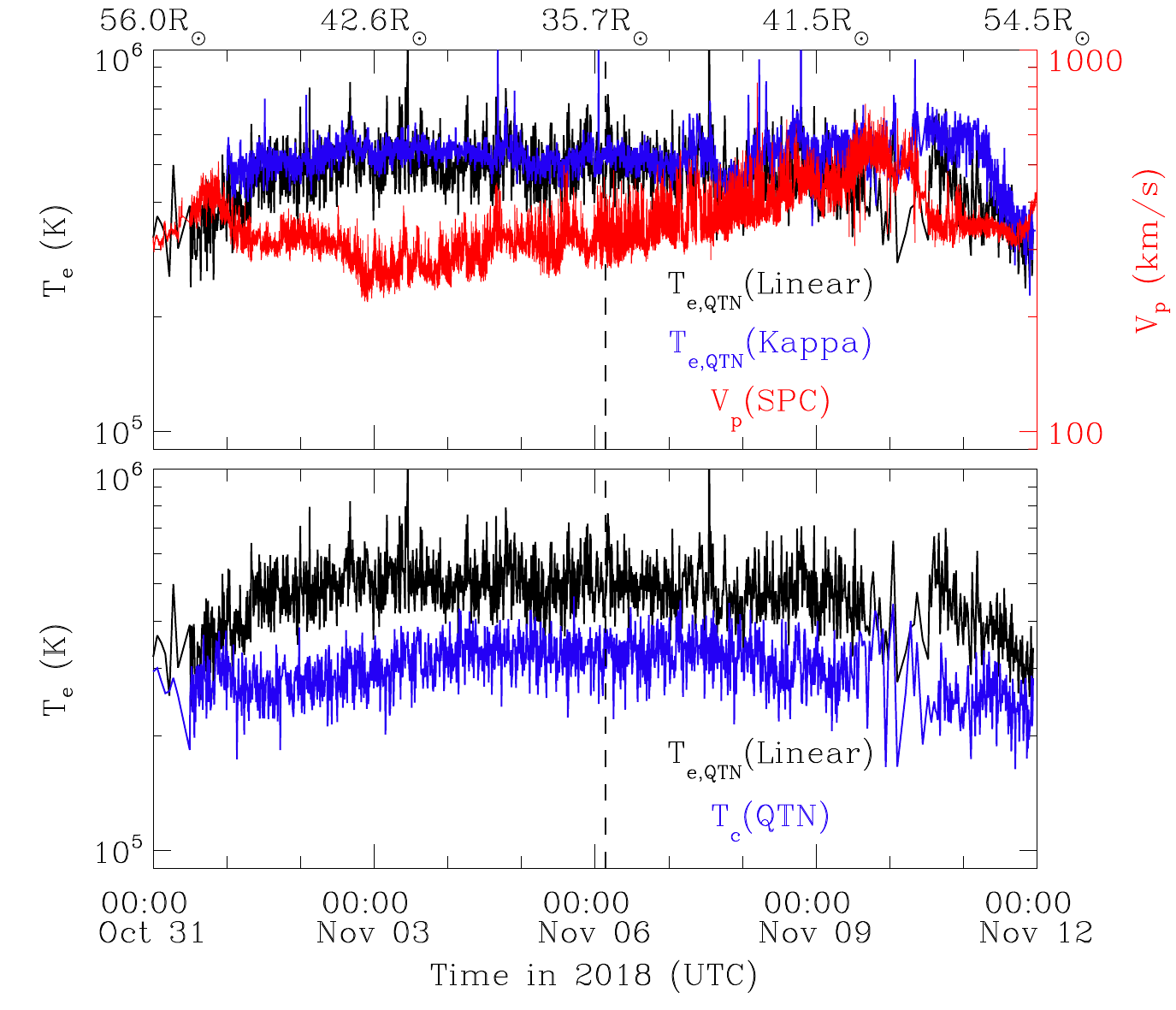}
	\caption{Comparison of observations of solar wind electron temperatures derived from different methods on Parker Solar Probe (PSP). An example of 12-day period of measurements by PSP during Encounter One (from October 31, 2018 00:00:00 to November 12, 2018 00:00:00 UTC) is shown for reference. The heliocentric distance (in units of the solar radius $R_{\sun}$) is indicated at the top of the top panel and the black vertical line denotes the first perihelion of the PSP orbit. From the top to bottom panels, the total electron temperature derived from the linear fit QTN technique is displayed in black. In the top panel, the total electron temperature obtained by fitting the high-frequency part of the spectrum with the generalized Lorentzian QTN model \citep{Maksimovic2020} is shown in blue for comparison. The bottom panel follows the same format as the first panel but for the core electron temperature derived from the simplified QTN technique \citep{2020Moncuquet}. The proton bulk speed from SPC/SWEAP is presented in the top panel for reference. An anticorrelation between $V_p$ and $T_e$, which was also previously reported in \cite{Maksimovic2020}, is visible during the time interval considered. Note that we have already smoothed $T_{e,QTN}$(Linear), $T_{c,QTN}$ and $T_{e,QTN}$(Kappa), so that the comparison between them is clear.}
	\label{f2}
\end{figure}

Figure~\ref{f2} shows an overview of the solar wind electron temperatures and the proton bulk speed measured by PSP during E01 (from October 31, 2018 to November 12, 2018 UTC). The electron temperatures derived from different techniques including $T_e$ from QTN (this study) and $T_c$ from QTN are compared for cross-checking. In the top panel, we present $T_e$ (in black, labelled as $T_{e,QTN}$(Linear)) derived from our linear fit QTN technique explained above and compare it to $T_e$ (in blue, $T_{e,QTN}$(Kappa)) derived from the generalized Lorentzian QTN model \citep{Maksimovic2020}. In general, they are in broad agreement with each other. Therefore, both the absolute values and variations of $T_{e,QTN}$(Linear) should be reliable. Similarly, $T_c$ from QTN \citep{2020Moncuquet} is displayed in blue in the bottom panel and are compared to $T_{e,QTN}$(Linear) (in black). The ratio $T_e/T_c$ reflects the contribution of suprathermal electrons and should not be a constant. The median value of the $T_{e,QTN}$(Linear)$/$$T_{c,QTN}$ is about 1.41, which is close to the median value of $T_{e,QTN}$(Kappa)$/$$T_{c,SPAN-E}$ ($\sim$1.47) \citep[see][]{Maksimovic2020}. $T_{c,SPAN-E}$ is the core electron temperature derived from SPAN-E \citep{2020Halekasb,2020Halekasa}.  Finally, the proton bulk speed from SPC/SWEAP is presented in the top panel for reference. The example time interval considered also shows an anticorrelation between $V_p$ and $T_e$, which was previously reported in \cite{Maksimovic2020}. We will further discuss this result in Section \ref{3.3}.

We note that the ratio of $T_{e}$$/$$T_{c}$ mentioned in this paper seems to disagree with that discussed by \cite{2020Halekasa}, especially near the perihelion (i.e. $\leq$ 0.2 AU). This may be due to a systematic discrepancy in measuring the suprathermal electrons between the QTN technique and the SPAN-E instrument. For the SPAN-E instrument, measurements of both halo and strahl electrons may have some caveats \cite[see][]{Whittlesey_2020,2020Halekasa,Maksimovic2021,2020Bercic}. These caveats combined make it more complicated to accuratly measure the total electron temperature by SPAN-E than the core electron temperature. For the QTN spectroscopy, uncertainties on the measurements are discussed in Section \ref{particle_distribution}. All these factors may at least partly contribute to the systematic difference, but they cannot explain the magnitude of the difference. Therefore, an accurate and detailed comparison of the QTN total electron temperature with the one measured by SPAN-E should be made with more care and needs further investigations. Similarly, an in-depth comparison between the core temperatures measured by the QTN \citep[e.g.,][]{2020Moncuquet} and SPAN-E \citep[e.g.,][]{2020Halekasa,2022Halekas} would also be useful, but is out of the scope of the present paper which is focused on the total electron temperature measurements from the high frequency part of the QTN spectra.

\section{Observations and Results} \label{3}

PSP was designed to gradually shrink its orbit around the Sun and get closer step by step via seven Venus gravity assist flybys within about seven years. In this work, we focus on the 12-day period of observations around each perihelion from E01 to E10 (E08 not included) with the heliocentric distance varying from about 13.0 to 60.0 $R_{\sun}$. During its first three encounters, PSP followed similar trajectories and reached the perihelion of 35.66 $R_{\sun}$ ($\sim$0.17 AU). In the following two orbits (from E04 to E05), PSP travelled closer to the Sun and reached perihelion of 27.8 $R_{\sun}$ ($\sim$0.13 AU). The perihelia of PSP orbits became about 20.8 $R_{\sun}$ for both E06 and E07 and about 16 $R_{\sun}$ for both E08 and E09. During E10, PSP reached as close to the Sun as 13 $R_{\sun}$. In Section \ref{3.1}, we provide an overview of the radial evolution of the total electron temperature derived from the QTN technique, combining the datasets from E01 to E10 (E08 not included). In Section \ref{3.2}, we analyze and discuss the electron temperature gradients for different solar wind populations classified by the proton bulk speed and the solar wind mass flux, respectively. In Section \ref{3.3}, we investigate the radial evolution of anticorrelation between $V_p$ and $T_e$.

\subsection{Mean Radial Profiles of $T_e$} \label{3.1}

\begin{figure}
	\centering
        \includegraphics[width=0.5\textwidth]{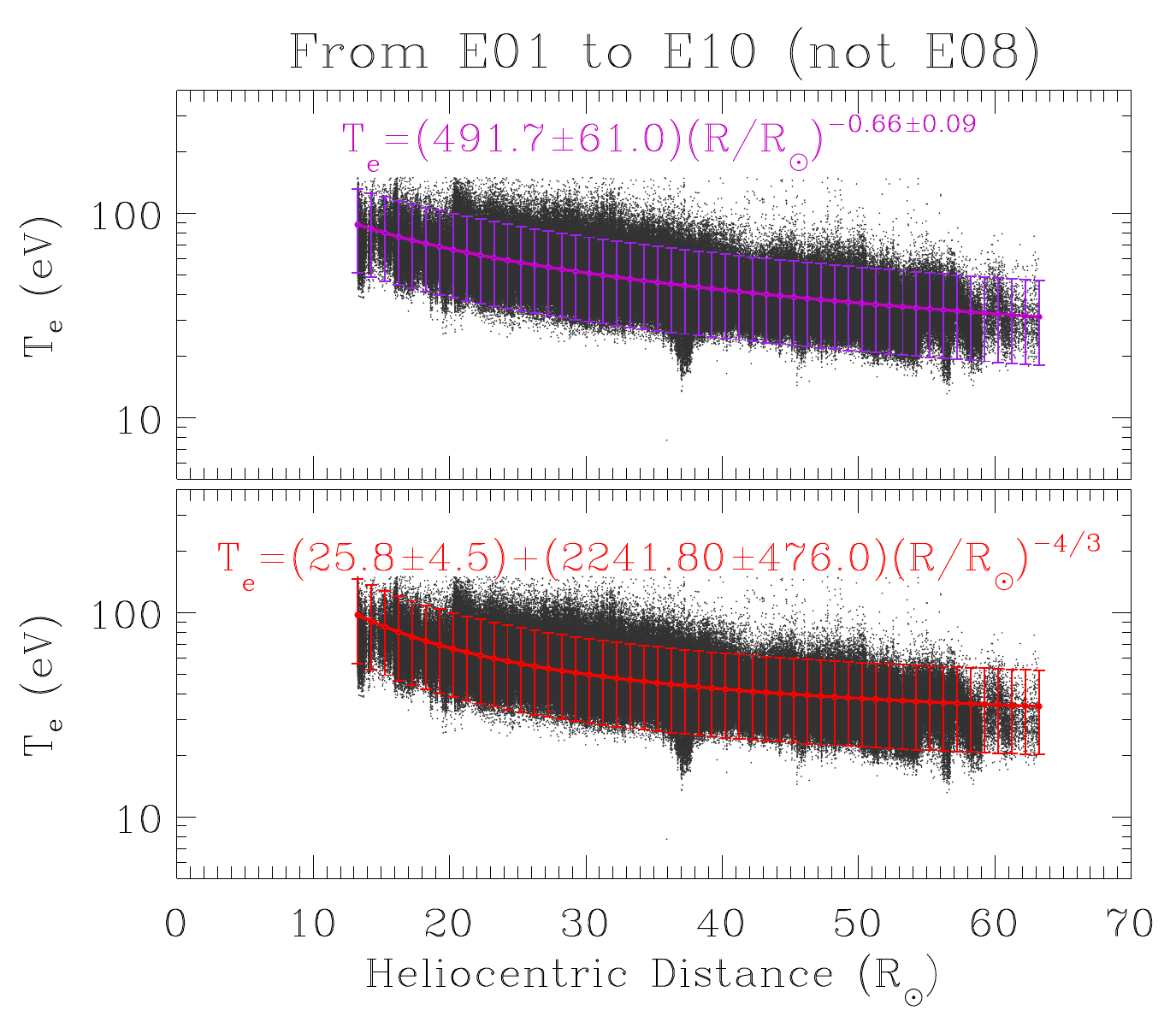}
	\caption{Radial variation of the total electron temperature ($T_e$) combining observations from Encounter One (E01) to Ten (E10) with E08 excluded. From top to bottom, $T_e$ was fitted with the power law expression $T_e = T_0\times(R/{R_{\sun}})^{\beta}$ (purple) and the expression given by the large distance exospheric solar wind model $T_e = T_0 +T_1\times(R/{R_{\sun}})^{-4/3}$ (red), respectively. The fitted profile and expression with corresponding color are superimposed for comparison. The variations of the free parameters in the fitted expressions are the corresponding 1-sigma (1-$\sigma$) fit uncertainties. The vertical error bars indicate the 2-$\sigma$ fit uncertainties, which in total covers about 95$\%$ data points.}
	\label{f3}
\end{figure}

Figure~\ref{f3} presents the total electron temperature derived from the QTN technique as a function of the heliocentric distance in units of solar radius $R_{\sun}$, combining 12-day period of observations near the perihelion of each encounter from E01 to E10 (E08 not included). Since PSP almost corotates with the Sun near the perihelion of each encounter, its observations only cover a very small heliographic latitude and longitude span \citep{2019Kasper, 2020Halekasb}. This means that, in each encounter, PSP detects the solar wind from only a limited number of sources. Therefore, a large data set from different encounters is necessary to remove/reduce the effects of transient structures such as CMEs or small-scale flux ropes \citep[e.g.,][]{2020Hess, 2020Zhao, 2020Korreck, 2021Chen}, switchbacks \citep[e.g.,][]{2019Bale,2020DudokdeWit,2021Martinovic,2021Fargette}, magnetic holes associated with slow shock pairs \citep[e.g.,][]{2021Chen,2022Zhou}, and so on. As explained below, we fit the total electron temperature with respect to the heliocentric distance with both the power law model and the exospheric model to get their mean radial profiles.

Specifically, we perform the fittings for each model by minimizing the $\chi^2$ value with the implementation of a nonlinear least-squares Levenberg-Marquardt technique \citep{2009Markwardt}. This technique takes into account the heliocentric distance and all the data points, as is generally the case for previous studies \citep[e.g.,][]{1998Issautier,2015JGRA,2013Hellinger,2020Moncuquet}. In total, there are $N \sim 882,361$ data points and there are two adjustable free parameters for each model fit. Therefore, the degree of freedom is $DOF=N-2=882,359$. $\chi^2$ is defined as $\chi^2=\sum_{i=1}^N (\frac{O_i-E_i}{\sigma_i})^2$, where $O_i$ is the value of the observations ($T_e$), $E_i$ is the corresponding expected value (fit), and $\sigma_i$ is the uncertainty of the measured $T_e$. As shown in section \ref{particle_distribution}, we estimate that $\sigma \sim 0.2 \times T_e$. The power-law model is derived with $\chi^2\simeq1178937$ and the so-called reduced/normalized $\chi^2_\nu=\chi^2/DOF\simeq1.34$. The exospheric model is derived with $\chi^2\simeq1235002$ and the so-called reduced/normalized $\chi^2_\nu=\chi^2/DOF\simeq1.40$. Since $\chi^2_\nu$ for both model fits are close to unity and are comparable in the two cases, one can conclude that the exospheric temperature model of the form $T_e = T_0+ T_1 \times r^{(-4/3)}$ is as good as the power law approximation in fitting the observed total electron temperature gradient in the small radial range considered. The fitted profiles and expressions for both models are shown on Figure~\ref{f3}. Furthermore, both the mean and median values of $\overline{T_e}/T_e$ are very close to unity for both model fits, where $\overline{T_e}$ is the fitted value and $T_e$ is the measured value. This again indicates the goodness of both model fittings. 1-$\sigma$ value of $\overline{T_e}/T_e$ for both model fits is around 0.2, based on which the uncertainties of the two free parameters for each model fit are derived. 2-$\sigma$ fit uncertainties are plotted in Figure~\ref{f3} for reference, which in total covers about 95$\%$ data points.

The total electron temperature fitted by the power law model ($T_e \propto r^{-0.66}$, where $r$ is the heliocentric distance in unit of solar radius) is displayed in purple. The derived total electron temperature profile is flatter than that of the core electron temperature \citep[$T_c \propto r^{-0.74}$, see][]{2020Moncuquet}, which is consistent with the results in the outer heliosphere \citep[e.g.,][]{1998Issautier,2011LeChat}. The total electron temperature consists of the contribution of the core, halo and strahl electron thermal pressure. Therefore, the flatter radial profile of $T_e$ may be explained by the nearly isothermal behaviour of suprathermal electrons \citep[see][]{2020Moncuquet}. We note that in that study, the suprathermal temperature is the total contribution of both the halo and strahl electron thermal pressures. Based on the SPAN-E observations \citep{2020Bercic}, there is no strong trend in variation of the strahl electron temperature with radial distance. Also, the strahl electrons are more pronounced closer to the Sun while the density ratio between the halo and strahl electrons increases with the radial distance \citep{2005Maksimovic,2009JGRA}, which suggests a conversion of some strahl electrons into halo ones. As a result, the fact that the $T_e$ profile is flatter than the $T_c$ one may mainly be due to the flatness of the strahl electron temperature profile. The recent results from PSP \citep[see][]{2022Abraham} suggest that the physical picture is somewhat different from a simple conversion of strahl to halo as discussed above. PSP results instead show that the overall suprathermal electron fraction (halo + strahl) increases with respect to the heliocentric distance below 0.25 AU, and that the halo and strahl relative density are quite small near perihelion. However, as is discussed in Section \ref{cross-checking}, close to the Sun, there are some caveats to measure both halo and strahl electrons by SPAN-E. The overall suprathermal fraction (halo + strahl) close to the Sun \citep[e.g.,][]{2022Abraham, Maksimovic2021} may be underestimated, both of which should be treated with more care.

\begin{figure*}
		\centering
		\begin{tabular}{cl}
		\includegraphics[width=0.33\textwidth]{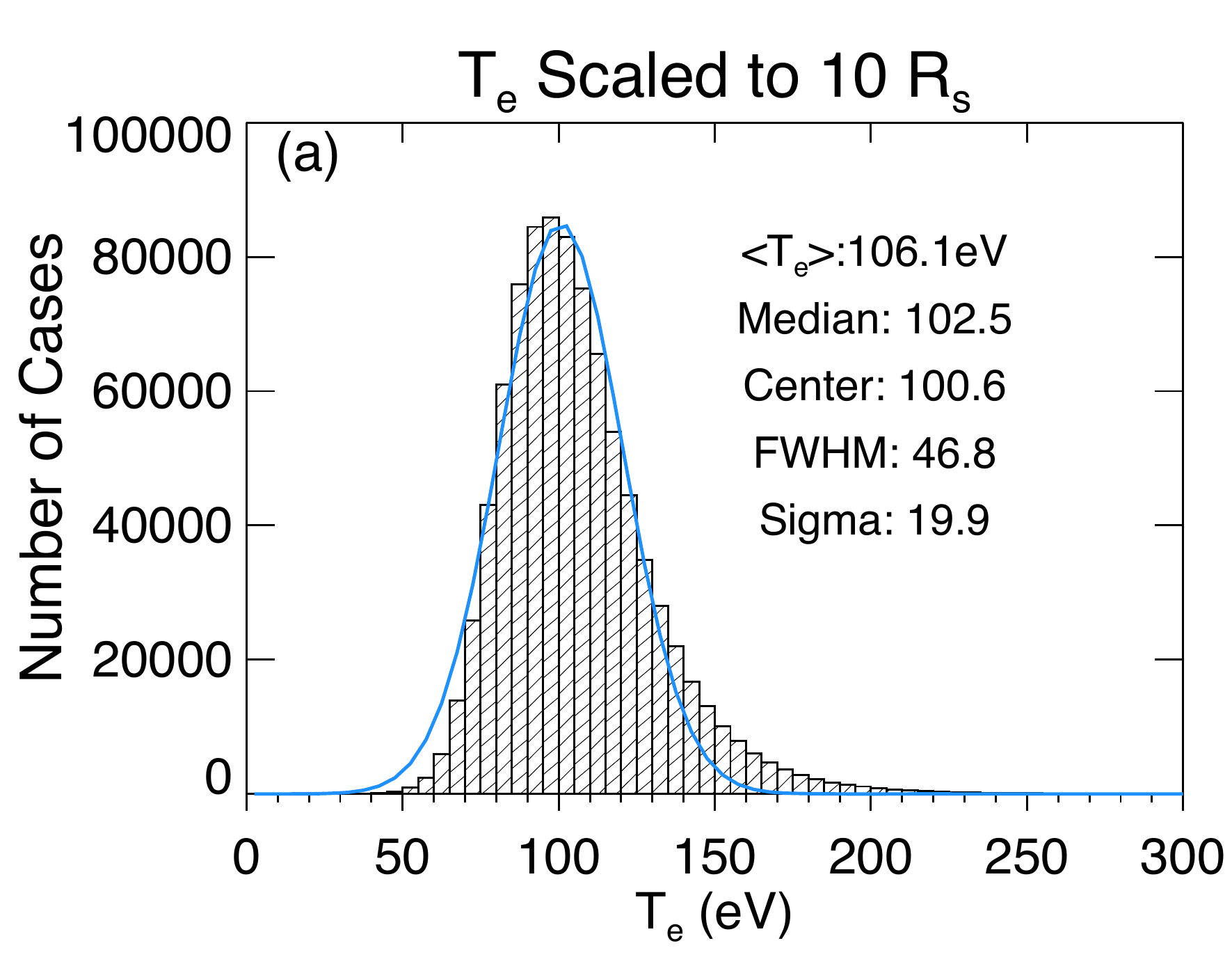}
		\includegraphics[width=0.33\textwidth]{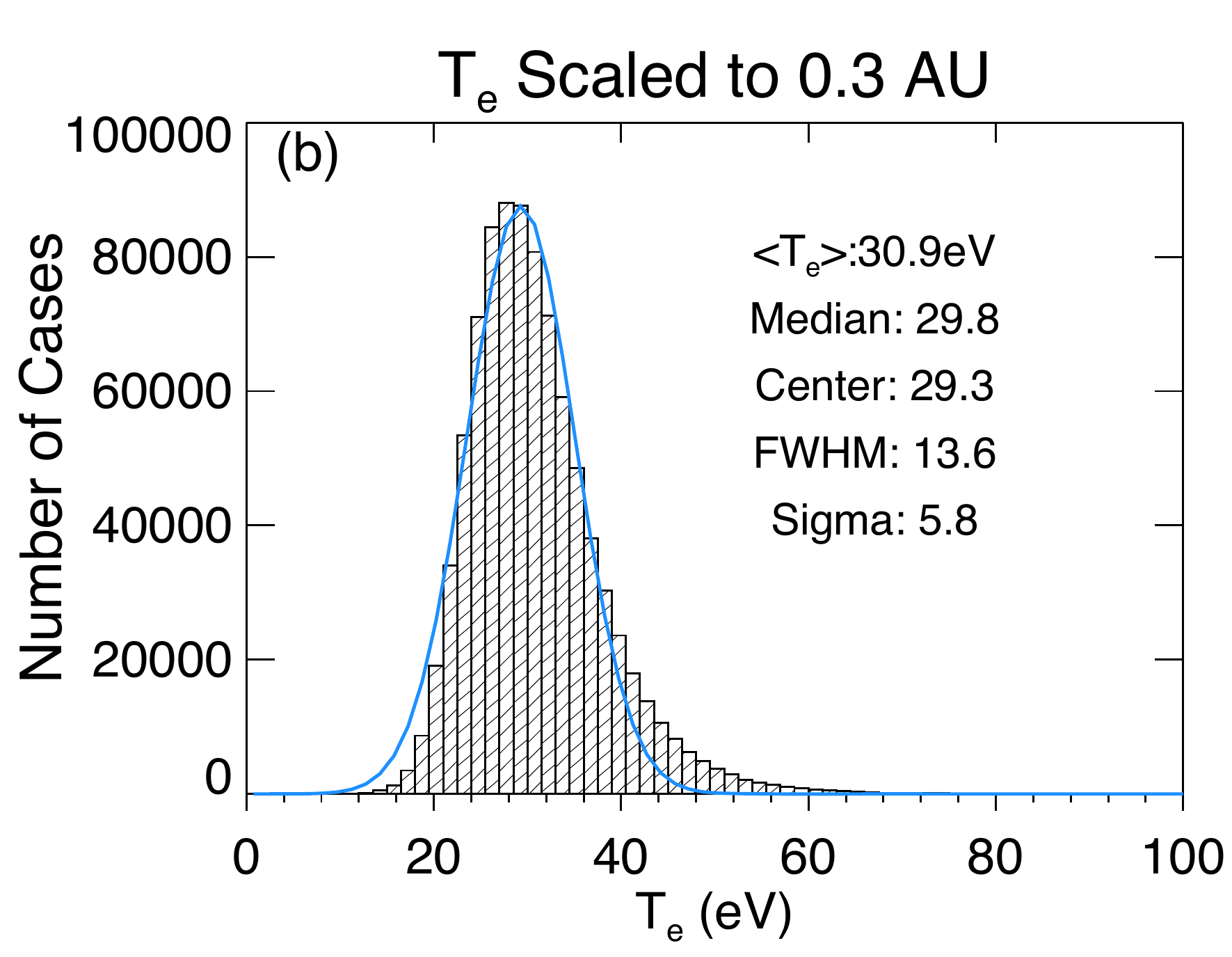}
		\includegraphics[width=0.33\textwidth]{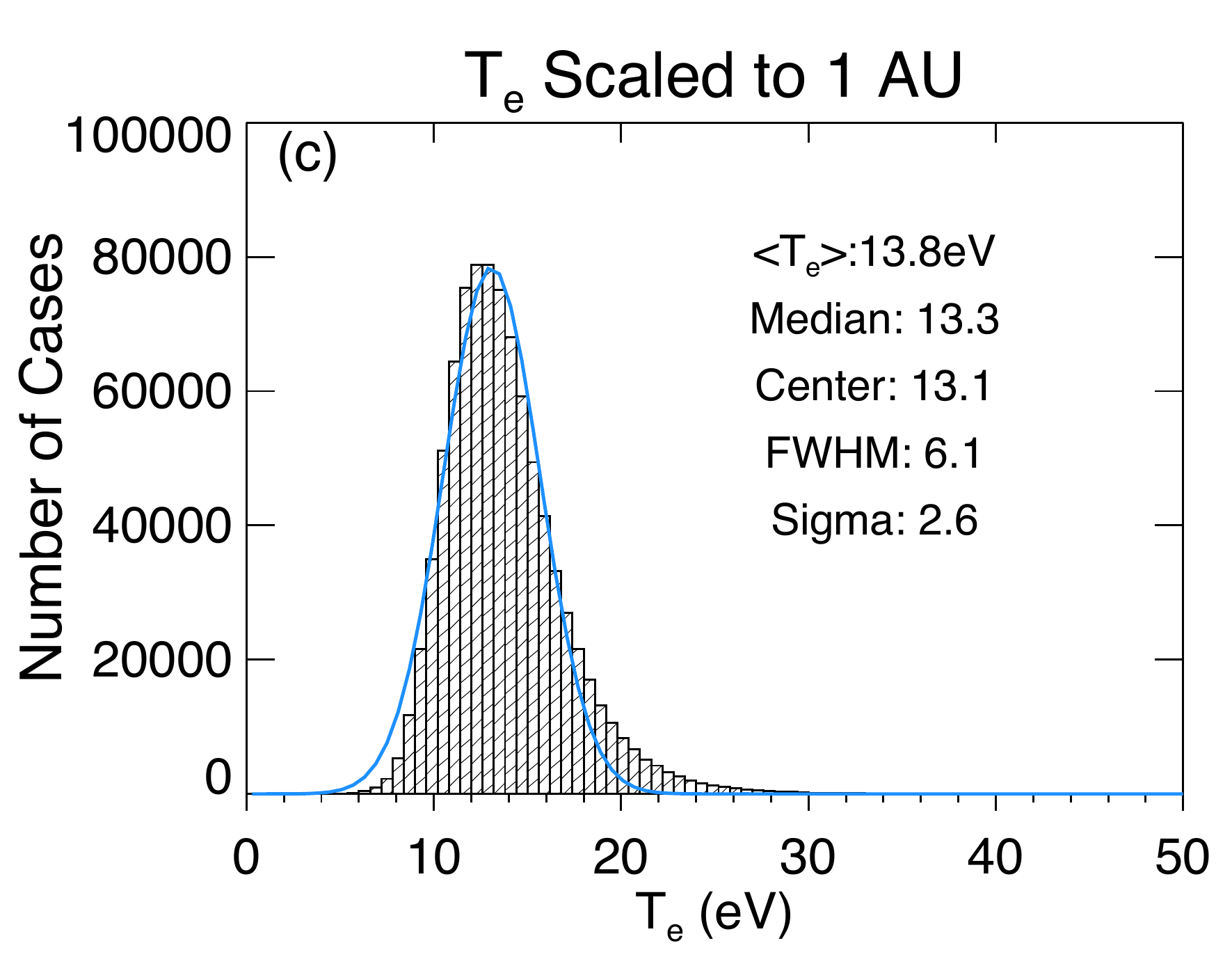}\\
		\end{tabular}
		\caption{(a--c): Histograms of the total electron temperature ($T_e$) scaled to 10 solar radii, 0.3 AU and 1 AU, based on observations displayed in Figure~\ref{f3} and their corresponding power law modelling fit results. Gaussian fit is superimposed in blue on each histogram. The corresponding average and median values are also indicated together with the center value and 1-$\sigma$ standard deviation of the Gaussian fit.}
		\label{f4}
\end{figure*}

Then, based on the power-law fitted $T_e$ profile ($T_e \propto r^{-0.66}$), we extrapolate $T_e$ to 10 $R_{\sun}$, 0.3 AU, and 1 AU, respectively. Figure~\ref{f4} (a), (b), and (c) show the corresponding distributions of the scaled $T_e$ combining the observations from E01 to E10 (E08 not included). A Gaussian function (blue line) was fitted on each histogram distribution and the corresponding center value (the most probable value) and 1-$\sigma$ standard deviation of Gaussian fit are shown in comparison with the mean and median values. The histogram distributions of $T_e$ are very symmetrical and almost Gaussian. Again, the difference between the mean, median, and the center value of Gaussian fit is quite small (less than 6$\%$). This may be explained by the fact that we combine observations from several different encounters (different types of wind from different sources). The exospheric solar wind model indicates that for $r$ $\textless$ 10 $R_{\sun}$, the $T_e$ radial profile becomes less steep \citep{2004Zouganelis}. So, when extrapolating $T_e$ back to the Sun with a constant slope, we stop the extrapolation at about 10 $R_{\sun}$. The value of $T_e$ scaled to 10 $R_{\sun}$ is around 100.6$\pm$19.9 eV. The predicted absolute values here are somewhat larger than the predictions shown in \cite{Bale2016}; however, they are similar to the strahl electron temperature measured by SPAN-E/SWEAP \citep{2020Bercic, Maksimovic2021}. The strahl electron temperature is considered to be closely related to or almost equal to the coronal electron temperature. At 10 $R_{\sun}$, this extrapolated temperature is also consistent with the exospheric solar wind model prediction derived from an electron velocity distribution with a Kappa index ranging between 4 and 6 \citep{2004Zouganelis}, which indicates that the electron distribution has a suprathermal tail as measured by the QTN measurements \citep[e.g.,][]{Maksimovic2020}. That same model yields a solar wind bulk speed between 250 and 350 km s$^{-1}$. Note that the Kappa index mentioned here is based on one unique generalized Lorentzian or Kappa function that is an alternative to the Maxwellian core plus Kappa/Maxwellian halo model. But the suprathermal tail itself may have a large kappa index, as found by SPAN-E near perihelion \citep[e.g.,][]{Maksimovic2021,2022Abraham}. Indeed, \cite{2004Zouganelis} showed that the acceleration provided by the exospheric model does not require specifically a Kappa function, but results more generally from nonthermal distributions. Our results show that the agreement between the extrapolated $T_e$ based on PSP observations and the exospheric solar wind model prediction is quite good, given the simplifications made in both the $T_e$ measurements and the solar wind model. Note that, $T_e$ scaled to 0.3 AU is $\sim$29.3$\pm$5.8 eV, which is consistent with the Helios observations at the same heliocentric distance \citep{2005Maksimovic}. For $T_e$ scaled to 1 AU, the value is $\sim$13.1$\pm$2.6 eV, which is almost the same as the long-term ($\sim$10 years) Wind observations \citep{2018Wilson}. $T_e$ scaled to 1 AU is also approximately the same as the mean/median value of the one-year statistical analysis based on STEREO observations \citep{2016Martinovic}. Note also that the extrapolated electron temperatures from the exospheric model fit (not shown here) are always higher than but still comparable to those from the power law model fit.

\subsection{Temperature gradients for different solar wind populations} \label{3.2}

\begin{figure*}
		\centering
		\begin{tabular}{cl}
		\includegraphics[width=0.45\textwidth]{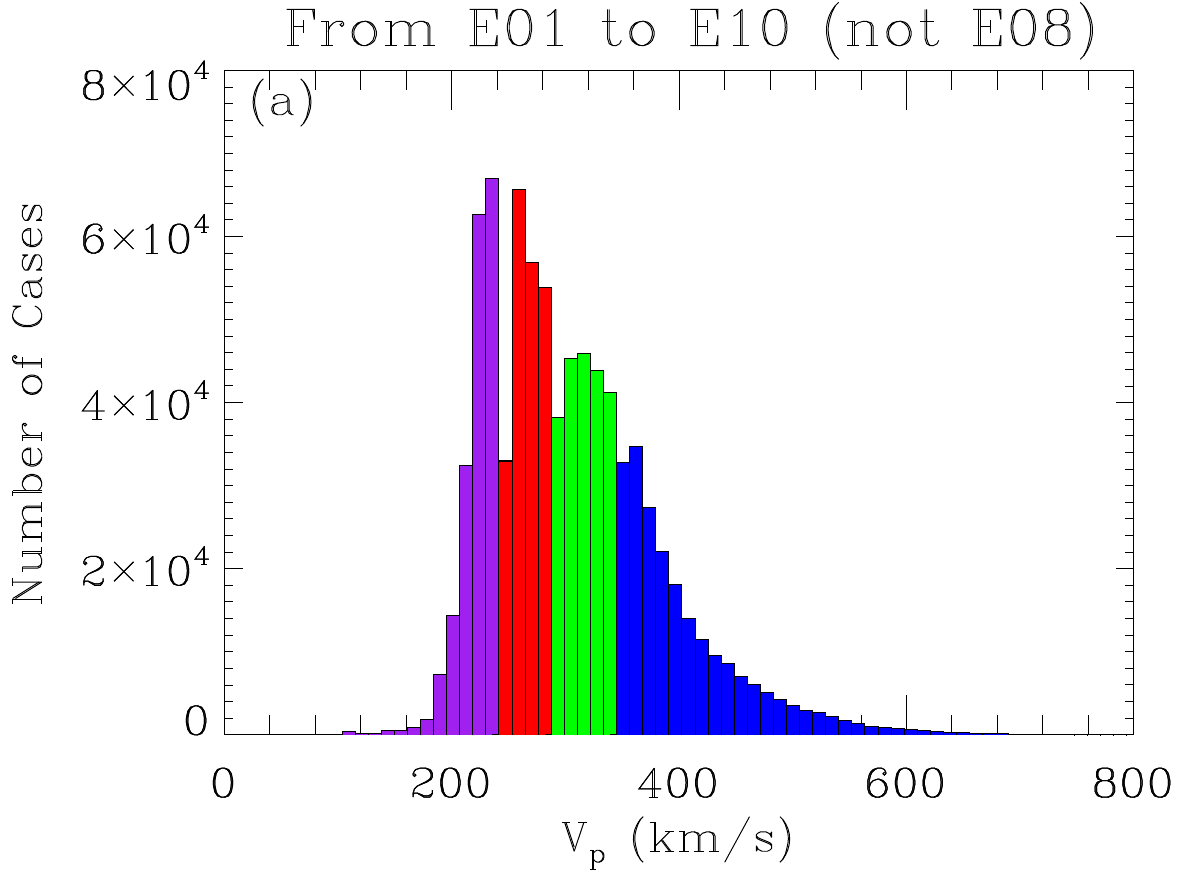}
		\includegraphics[width=0.45\textwidth]{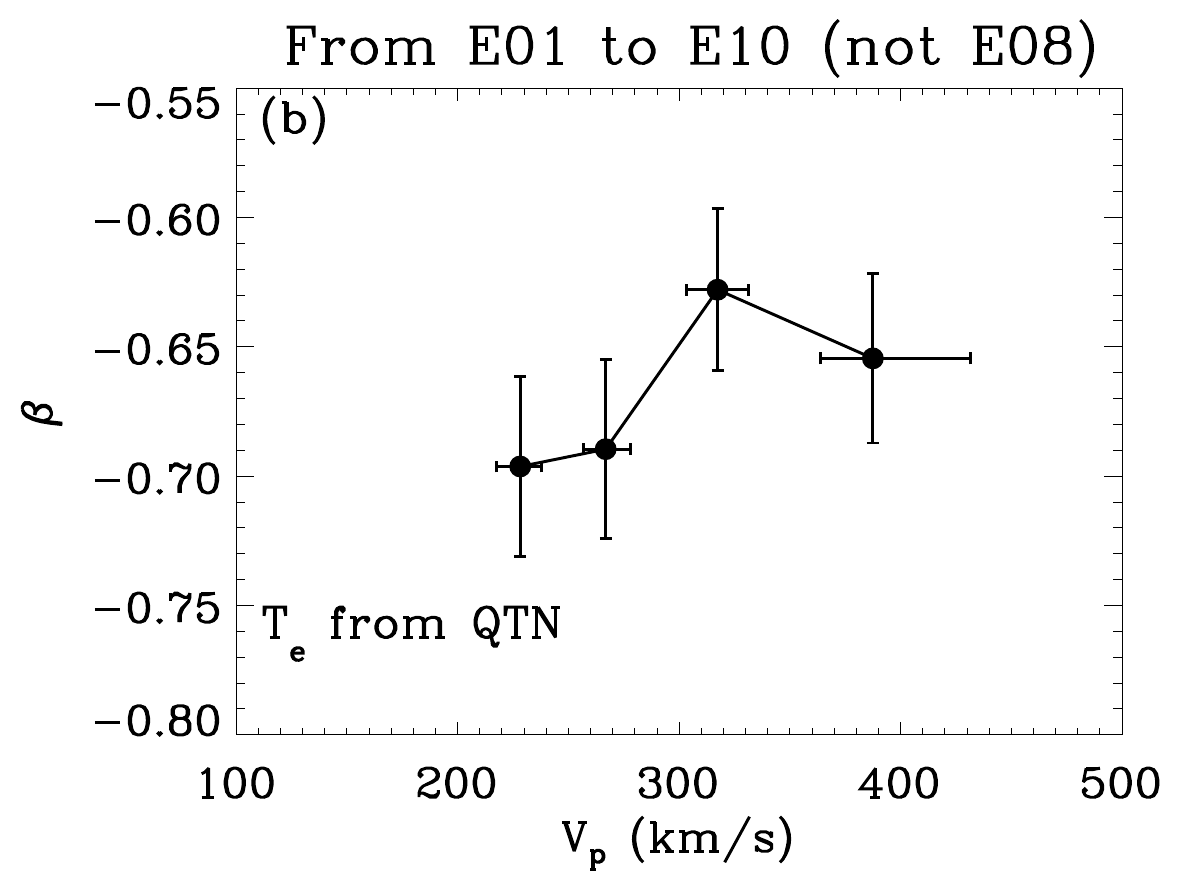}\\
		\end{tabular}
		\caption{(a) We define the four wind families based on the proton bulk speed histogram. Each colored histogram has the same number of observations. (b) Outcome of the power law modelling in the form $T_e = T_0\times(R/{R_{\sun}})^{\beta}$ for total electron temperature: $\beta$ versus $V_p$. More details are described in the main text.}
		\label{f5}
\end{figure*}

As presented/discussed in previous investigations \citep[e.g.,][]{1998meyervernet,2011LeChat,2005Maksimovic,2009JGRA,2015JGRA,Maksimovic2020}, solar wind classified based on the proton bulk speed may have different electron heating and cooling behaviours. Therefore, in order to do direct comparisons with the previous studies, we also separate solar wind populations based on the proton bulk speed. The dataset was split into four proton bulk speed bins as illustrated by Figure~\ref{f5} (a). In this way, each proton bulk speed bin contains the same number of data points, which is 882,361/4 $\sim$ 220590. We used the total proton bulk speed ($V_p$) provided by SPC/SWEAP for E01 and E02 and those from SPAN-I/SWEAP after E02 \citep{Kasper2016, Case2020}. For each proton bulk speed bin, we fit the $T_e$ radial profile with a power-law model using the method described in section \ref{3.1}. The derived power law indices are plotted against the corresponding proton bulk speed in Figure~\ref{f5} (b). We use the proton bulk velocity measured in the RTN coordinate system. The radial component of the velocity ($V_R$) measured by SPC and SPAN-I are in good agreement, but there is a systematic disprepancy for the tangential component ($V_T$) \citep{2021Woodham}. However, $V_R$ is the main component of $V_p$ (total proton bulk speed), and their absolute values are very close to each other. Furthermore, we use both $V_R$ and $V_p$ to cross-check the results below in this section and in section \ref{3.3}. We verify that the measurement uncertainty of $V_p$ does not affect our conclusions.

The $T_e$ radial gradients have a tendency (though weak) for the slower wind electrons to cool down with a steeper profile than the faster wind ones. It is noteworthy to mention that with only 12-day period of observations for each encounter (from E01 to E10, with E08 excluded) and a limited latitude exploration, we find similar behaviour for electrons in the inner heliosphere as previous long-term investigations \citep[e.g.,][]{2005Maksimovic,2015JGRA,Maksimovic2020} at various latitudes and longitudes and much larger span of heliocentric distances in the outer heliosphere. This is also consistent with the exospheric model predictions as shown in \cite{1998meyervernet}. Also, we note that the $T_e$ radial gradient within each proton bulk speed bin is steeper than that in the outer heliosphere based on Ulysses observations ($T_e \propto r^{-0.53}$, see \cite{2011LeChat}). This may verify the exospheric model prediction that the electron temperature profile becomes steeper when getting closer to the Sun \citep{1998meyervernet}. 

\begin{figure*}
		\centering
		\begin{tabular}{cl}
		\includegraphics[width=0.45\textwidth]{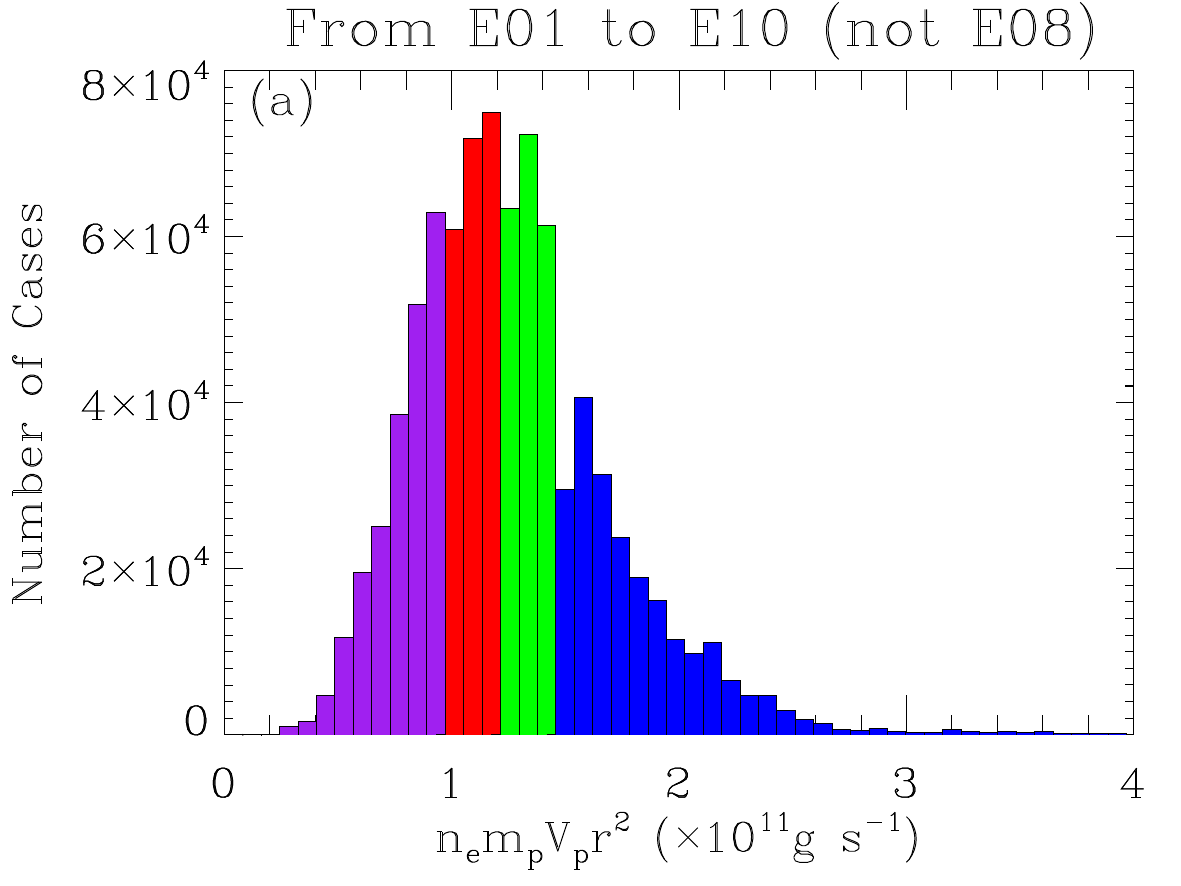}
		\includegraphics[width=0.45\textwidth]{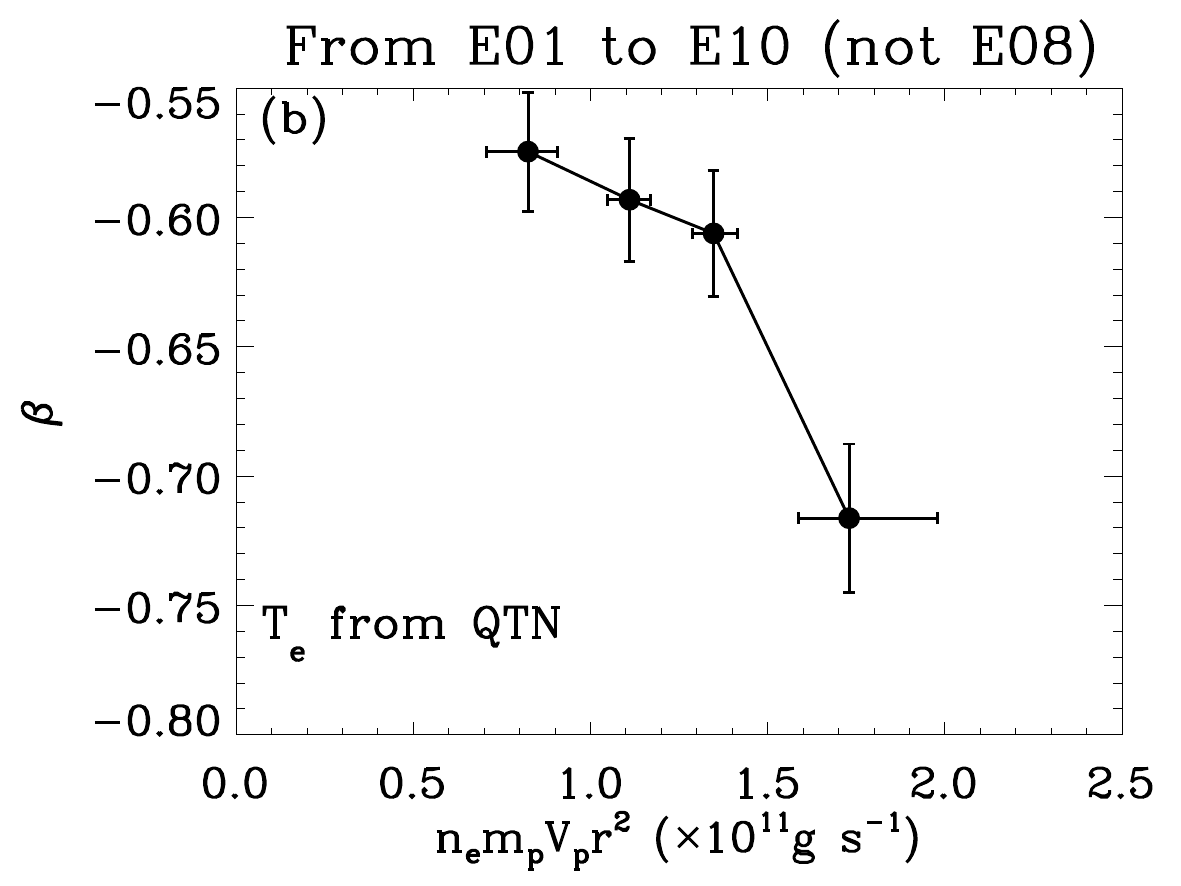}\\
		\end{tabular}
		\caption{Follows the same format as Figure~\ref{f5} but for the solar wind mass flux.}
		\label{f6}
\end{figure*}

Moreover, since PSP is very close to the Sun during the encounter phase where the solar wind is still under acceleration, the proton bulk speed detected by PSP may not be the final speed. Therefore, because of the different types of winds coming from different source regions, we further use another basic physical quantity to partition the dataset based on almost constant streamline, i.e., the solar wind mass flux $F_w=n_em_pV_pr^2$ \citep[e.g.,][]{1990Wang,2017Bemporad}. The resulting histogram distribution of $F_w$ is shown in Figure~\ref{f6} (a). The derived values ($F_w \sim 2 \times 10^{10} - 3 \times 10^{11}$g s$^{-1}$) are in agreement with the remote--sensing observations from SOHO at altitudes higher than 3.5 $R_{\sun}$ \citep{2017Bemporad}, in situ measurements from ACE at 1 AU \citep{2010Wang}, and in situ data by Ulysses from $\sim$1.4 to $\sim$1.8 AU \citep{2008Issautier,2010Wang}. As expected, this indicates the conservation of the solar wind mass flux. \cite{2010Wang} showed that the solar wind mass flux at the corona base increases roughly with the footpoint field strength. This indicates, to some degree, both the corona base conditions and the propagation effects are considered for $F_w$, in contrast to the proton bulk speed. Thus, as displayed in Figure~\ref{f6} (a), we split the dataset into four solar wind mass flux tubes and check the corresponding electron temperature gradients. Figure~\ref{f6} (b) shows that solar wind electrons within the flux tube with larger mass flux cool down faster.

\subsection{Anticorrelated parameters: $V_p$ and $T_e$} \label{3.3}

As shown in section \ref{cross-checking}, PSP observations display a clear anticorrelation between $V_p$ and $T_e$ during E01. A similar anticorrelaton was observed during E04, E05, E07 and E09. During E02 and E10, frequent Type III radio emissions were detected by PSP and fewer effective data points of $T_e$ derived from the QTN technique were obtained than during other encounters. This may affect the analysis of the relation between $V_p$ and $T_e$. In contrast, slight correlated ($V_p$, $T_e$) were observed during E03 and E06 based on the QTN observations. The $V_p$--$T_e$ relation measured in the solar wind may indeed depend on both the source region \citep{2021Griton} and the radial evolution \citep{Maksimovic2020,2020Pierrard,2022Halekas}. The complexity of the electron temperature behaviours, especially the anticorrelation between $V_p$ and $T_e$, contrasts with the correlation between the proton temperature and the wind speed that persists throughout the heliosphere \citep[see][and references therein]{Maksimovic2020}. We selected the data points from E01, E04, E05, E07 and E09, and further analyzed the effect of the radial evolution on the anticorrelation between $V_p$ and $T_e$. 

\begin{figure*}
		\centering
		\begin{tabular}{cl}
		\includegraphics[width=0.3333\textwidth]{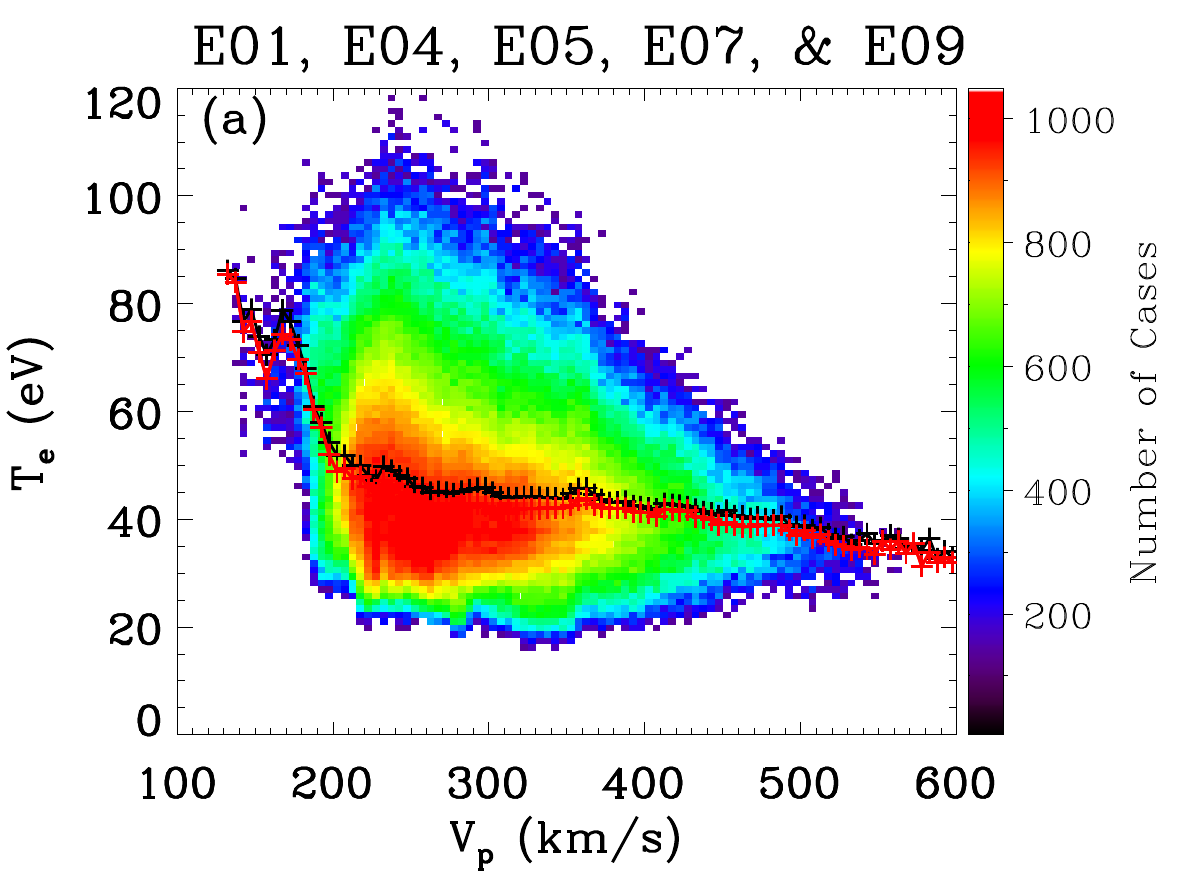}
		\includegraphics[width=0.3333\textwidth]{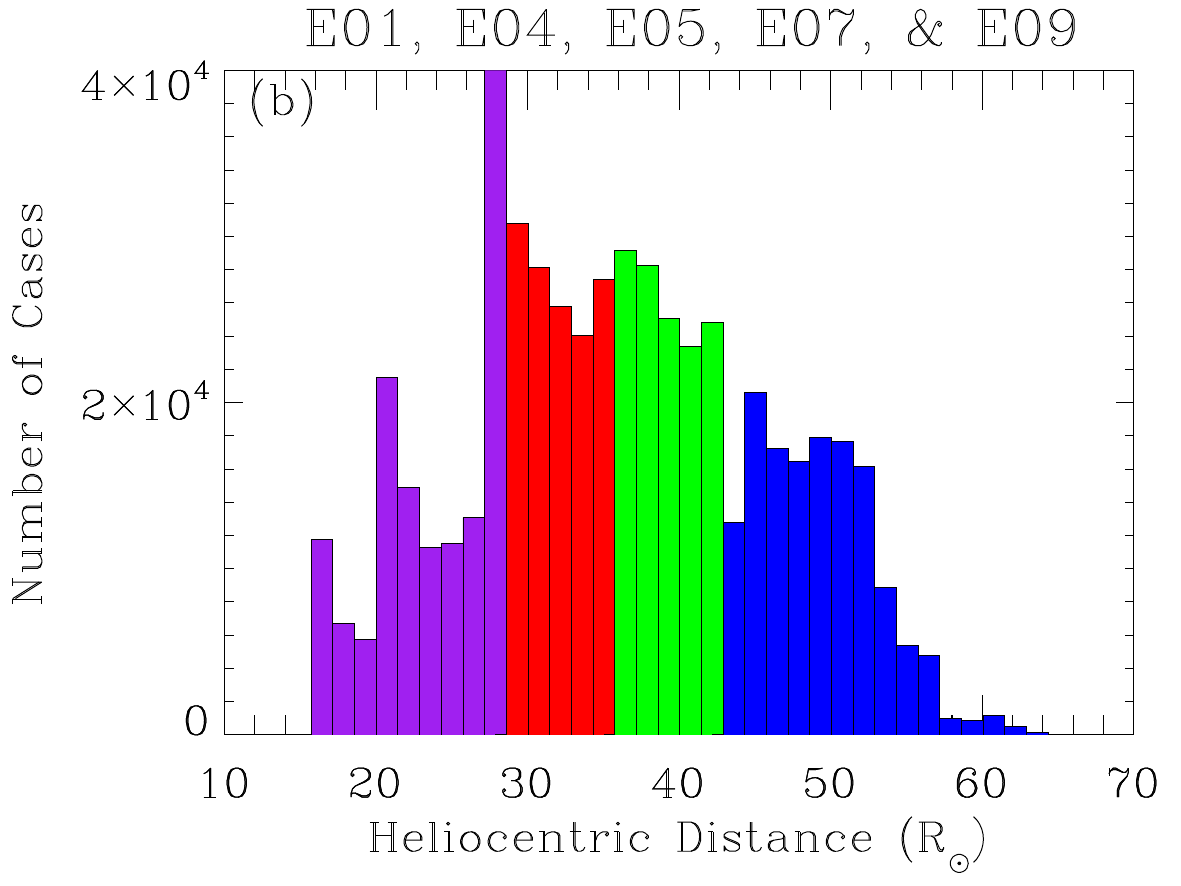}
		\includegraphics[width=0.3333\textwidth]{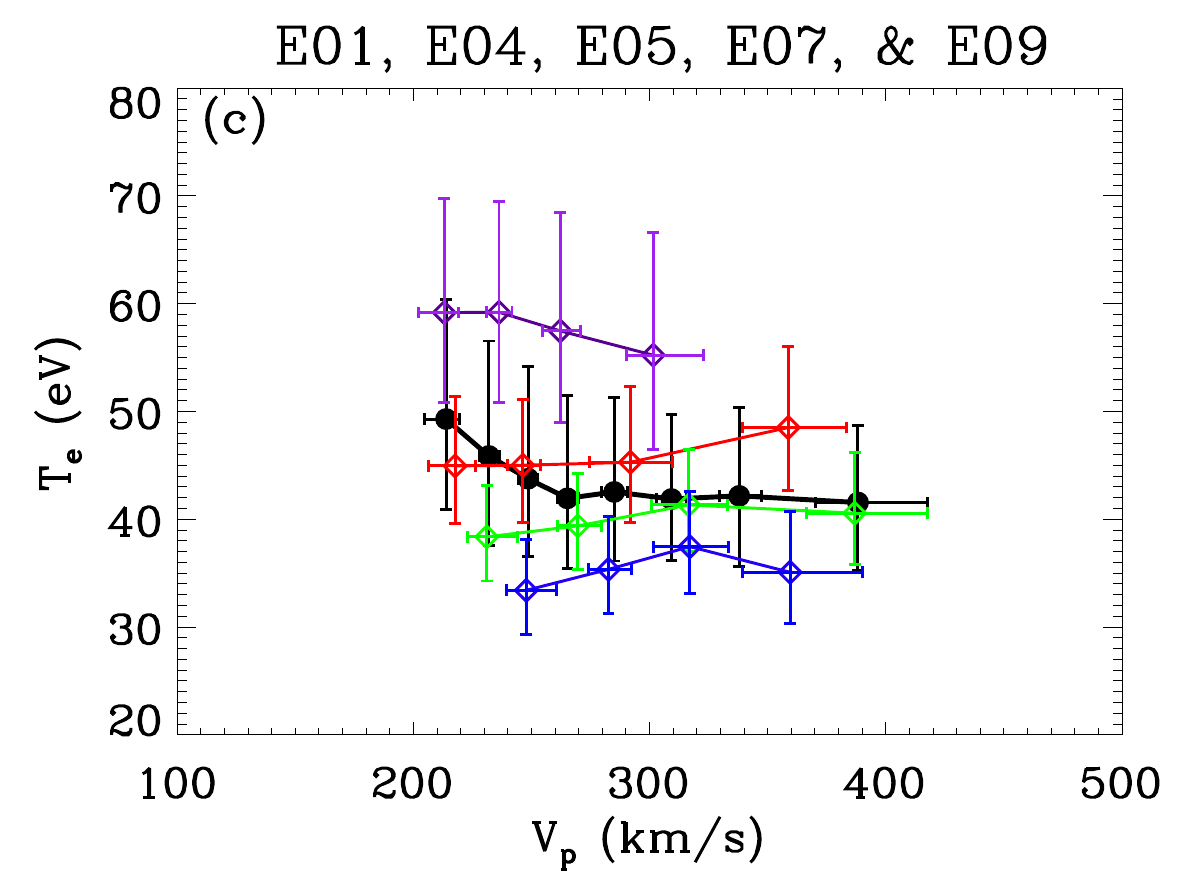}\\
		\end{tabular}
		\caption{(a) 2-D histogram ditribution of $T_e$ versus $V_p$ with the color bar on the right side indicating the number of data points. Both the mean (red curve) and median values (black curve) are superimposed for reference. A clear anticorrelated ($V_p$, $T_e$) is displayed. (b) We define the four wind families based on the heliocentric distance histogram. Each colored histogram has the same number of observations. (c) Relation between $V_p$ and $T_e$ for each wind family, as defined in panel (b). The results are displayed in the same color as the corresponding heliocentric distance histogram in panel (b). More details are described in the main text.}
		\label{f7}
\end{figure*}

Figure~\ref{f7} (a) compares $T_e$ and $V_p$ combining observations from E01, E04, E05, E07 and E09. A clear anticorrelated ($V_p$, $T_e$) is displayed. We also equally split the dataset into four radial distance bins as illustrated by Figure~\ref{f7} (b). For each radial distance bin, the datasets are equally split into four proton bulk speed bins following the method mentioned in section \ref{3.2}. We then compute the median values of $V_p$ and $T_e$ for each proton bulk speed bin. The calculated median values of $V_p$ and $T_e$ belonging to each radial distance bin are presented in the same color in Figure~\ref{f7} (c). For comparison, median values of $V_p$ and $T_e$ of the whole dataset equally divided into eight proton bulk speed bins are plotted in black.

We find that the ($V_p$, $T_e$) anticorrelation is stronger when the solar wind is slower (see black curve in Figure~\ref{f7} (c)). For the solar wind considered, most of them are slow wind and on average they are being accelerated during the expansion. Therefore, the slower solar wind is detected closer to the Sun. This is consistent with the fact that the most pronounced anticorrelated $V_p$--$T_e$ is observed close to the Sun (see purple curve in Figure~\ref{f7} (c)). The results may also indicate that the ($V_p$, $T_e$) anticorrelation is reduced/removed during the acceleration process of the slow solar wind. Based on both the Helios and PSP measurements, \cite{2020Bercic} found a clear anticorrelation between the parallel strahl electron temperature $T_{s\parallel}$ (proxy coronal electron temperature) and the local solar wind speed. \cite{2022Halekas} grouped the PSP observations by the asymptotic wind speed, and found that both the \insitu{} electron temperature (parallel core electron temperature $T_{c\parallel}$) and the proxy coronal electron temperature ($T_{s\parallel}$) are anticorrelated with the asymptotic wind speed. As a result, the anticorrelated ($V_p$, $T_e$) herein may be the remnants of the coronal conditions.

\section{Summary and Discussion} \label{summary_and_discussion}

In this work, we have implemented a simple, fast and effective method, based on the QTN spectroscopy, on PSP observations to derive the total electron temperature. To do so, we used the linear fit of the high frequency part of the QTN spectra observed by RFS/FIELDS. The derived total electron temperature is in broad agreement with $T_e$ obtained from the QTN model with Lorentzian velocity distribution functions \citep{Maksimovic2020}. We present the radial evolution of the total electron temperature by combining 12-day period of observations around each perihelion from E01 to E10 (E08 not included) with the heliocentric distance ranging from about 13 to 60 $R_{\sun}$.

The radial profile of the total electron temperature ($T_e \propto r^{-0.66}$) in the inner heliosphere falls within the range between adiabatic and isothermal and is flatter than that of the electron core temperature \citep[$T_c \propto r^{-0.74}$, see][]{2020Moncuquet}. This is consistent with previous Helios and Ulysses observations farther out \citep[e.g.,][]{1990Pilipp,1998Issautier,2011LeChat}. The flatness of the radial profile of $T_e$ may mainly be due to the contribution of the strahl electrons. The extrapolated $T_e$ to 0.3 AU and 1 AU using the fitted power law are almost the same as the Helios and Wind observations at the same heliocentric distance \citep[see][]{2005Maksimovic, 2018Wilson}, respectively. The total electron temperature extrapolated back to 10 $R_{\sun}$ is almost the same as the solar corona strahl electron temperature \citep{2020Bercic}. This may confirm that the strahl electron temperature is closely related to or even almost equals to the coronal electron temperature. The temperature extrapolated back to 10 $R_{\sun}$ is also consistent with the exospheric solar wind model prediction assuming an electron velocity distribution with the Kappa index ranging between 4 and 6 \citep{2004Zouganelis}. The extrapolated $T_e$ based on the exospheric solar wind model is systematically higher (but still comparable to) than that derived from the power-law model fit.

The radial $T_e$ profiles in the slower solar wind are relatively steeper than those in the faster solar wind. Stated in another way, electrons in the slower solar wind cool down more quickly than those in the faster wind. It is remarkable that with only 12-day period of observations for each encounter (from E01 to E10 with E08 excluded) and a limited latitude exploration, we find the same conclusions about electron cooling and heating behaviours in the inner heliosphere as previous long-term investigations \citep[e.g.,][]{2005Maksimovic,2015JGRA,Maksimovic2020} with a much larger span of the latitude, longitude and heliocentric distance explorations in the outer heliosphere. In general, the radial $T_e$ profile observed in the pristine solar wind is steeper than that in the outer heliosphere, which to some degree verify the exospheric model prediction in the inner heliosphere \citep{1998meyervernet}. This indicates that the exospheric solar wind model explains partially the electron behaviours in the inner heliosphere. Furthermore, the solar wind mass flux derived from the in situ dataset in the inner heliosphere from PSP are in agreement with those even closer to the corona base \citep{2017Bemporad} and further out in the interplanetary space \citep{2008Issautier,2010Wang}. Interestingly, the electron temperature cools down faster within the flux tube with larger mass flux. Given the solar wind mass flux conservation and the fact that the mass flux at the corona base increases with the footpoint field strength \citep{2010Wang}, it can be used as an effective physical quantity to distinguish the solar wind into different populations. This considers both the corona base conditions and the propagation effects in contrast to the proton bulk speed. Especially, the solar wind at distances of PSP orbit perihelia may be still accelerating, the speed should not be considered as the final speed.

With PSP observations from E01, E04, E05, E07 and E09, we find that the ($V_p$, $T_e$) anticorrelation is more pronounced when the solar wind is slower in the inner heliosphere. During the time period considered, most of the detected solar wind is slow wind, which on average is still being accelerated during the spherical expansion. Furthermore, the results may indicate that the slow solar wind acceleration during the expansion reduces/removes the strong ($V_p$, $T_e$) anticorrelation detected near the Sun. This is verified by the fact that the most pronounced anticorrelated $V_p$--$T_e$ is observed close to the Sun, in agreement with \cite{Maksimovic2020}. The solar wind $V_p$--$T_e$ relation is still an interesting issue, which may depend on both the source region in the Sun \citep{2021Griton} and the radial evolution during the expansion \citep{Maksimovic2020,2020Pierrard}. To comprehensively understand the $V_p$--$T_e$ relation, more work is needed to distinguish and/or connect the effects from the source region, spherical expansion and the transient structures detected locally.

Based on the Helios and preliminary PSP observations \citep[e.g.,][]{2005Maksimovic,2009JGRA,2020Halekasb,2020Halekasa,2020Bercic}, the strahl electrons will become more pronounced when PSP gets closer to the Sun. However, the QTN technique currently cannot resolve the strahl electrons well, which needs further theoretical/model extensions. When PSP gets closer to the Sun, $L/L_D$ is expected to become larger. This will enable us to derive the electron properties (e.g., $n_e$, $T_e$, and kappa index) with smaller uncertainties by fitting the whole QTN spectrum with the generalized Lorentzian (or so-called kappa) QTN model. Also, we await for well calibrated fixed $T_e$ from SPAN-E for all encounters to make systematic cross-checking with $T_e$ provided by the QTN technique, which should benefit to both methods.

\begin{acknowledgements}
			The research was supported by the CNES and DIM ACAV+ PhD funding. Parker Solar Probe was designed, built, and is now operated by the Johns Hopkins Applied Physics Laboratory as part of NASA’s Living with a Star (LWS) program (contract NNN06AA01C). Support from the LWS management and technical team has played a critical role in the success of the Parker Solar Probe mission. We acknowledge the use of data from FIELDS/PSP (http://research.ssl.berkeley.edu/data/psp/data/sci/fields/l2/) and SWEAP/PSP (http://sweap.cfa.harvard.edu/pub/data/sci/sweap/). We thank the CDPP (Centre de Données de la Physique des plasmas) and CNES (Centre National d'Etudes Spatiales) for their archiving and provision of SQTN data (https://cdpp-archive.cnes.fr/).
\end{acknowledgements}

%
%

\bibliography{bibliography}

\begin{thebibliography}{88}
\expandafter\ifx\csname natexlab\endcsname\relax\def\natexlab#1{#1}\fi

\bibitem[{{Abraham} {et~al.}(2022){Abraham}, {Owen}, {Verscharen}, {Bakrania},
  {Stansby}, {Wicks}, {Nicolaou}, {Whittlesey}, {Agudelo Rueda}, {Jeong}, \&
  {Ber{\v{c}}i{\v{c}}}}]{2022Abraham}
{Abraham}, J.~B., {Owen}, C.~J., {Verscharen}, D., {et~al.} 2022, \apj, 931,
  118

\bibitem[{{Bale} {et~al.}(2019){Bale}, {Badman}, {Bonnell}, {Bowen}, {Burgess},
  {Case}, {Cattell}, {Chandran}, {Chaston}, {Chen}, {Drake}, {de Wit},
  {Eastwood}, {Ergun}, {Farrell}, {Fong}, {Goetz}, {Goldstein}, {Goodrich},
  {Harvey}, {Horbury}, {Howes}, {Kasper}, {Kellogg}, {Klimchuk}, {Korreck},
  {Krasnoselskikh}, {Krucker}, {Laker}, {Larson}, {MacDowall}, {Maksimovic},
  {Malaspina}, {Martinez-Oliveros}, {McComas}, {Meyer-Vernet}, {Moncuquet},
  {Mozer}, {Phan}, {Pulupa}, {Raouafi}, {Salem}, {Stansby}, {Stevens}, {Szabo},
  {Velli}, {Woolley}, \& {Wygant}}]{2019Bale}
{Bale}, S.~D., {Badman}, S.~T., {Bonnell}, J.~W., {et~al.} 2019, \nat, 576, 237

\bibitem[{{Bale} {et~al.}(2016){Bale}, {Goetz}, {Harvey}, {Turin}, {Bonnell},
  {Dudok de Wit}, {Ergun}, {MacDowall}, {Pulupa}, {Andre}, {Bolton},
  {Bougeret}, {Bowen}, {Burgess}, {Cattell}, {Chandran}, {Chaston}, {Chen},
  {Choi}, {Connerney}, {Cranmer}, {Diaz-Aguado}, {Donakowski}, {Drake},
  {Farrell}, {Fergeau}, {Fermin}, {Fischer}, {Fox}, {Glaser}, {Goldstein},
  {Gordon}, {Hanson}, {Harris}, {Hayes}, {Hinze}, {Hollweg}, {Horbury},
  {Howard}, {Hoxie}, {Jannet}, {Karlsson}, {Kasper}, {Kellogg}, {Kien},
  {Klimchuk}, {Krasnoselskikh}, {Krucker}, {Lynch}, {Maksimovic}, {Malaspina},
  {Marker}, {Martin}, {Martinez-Oliveros}, {McCauley}, {McComas}, {McDonald},
  {Meyer-Vernet}, {Moncuquet}, {Monson}, {Mozer}, {Murphy}, {Odom},
  {Oliverson}, {Olson}, {Parker}, {Pankow}, {Phan}, {Quataert}, {Quinn},
  {Ruplin}, {Salem}, {Seitz}, {Sheppard}, {Siy}, {Stevens}, {Summers}, {Szabo},
  {Timofeeva}, {Vaivads}, {Velli}, {Yehle}, {Werthimer}, \&
  {Wygant}}]{Bale2016}
{Bale}, S.~D., {Goetz}, K., {Harvey}, P.~R., {et~al.} 2016, \ssr, 204, 49

\bibitem[{{Bemporad}(2017)}]{2017Bemporad}
{Bemporad}, A. 2017, \apj, 846, 86

\bibitem[{{Ber{\v{c}}i{\v{c}}} {et~al.}(2020){Ber{\v{c}}i{\v{c}}}, {Larson},
  {Whittlesey}, {Maksimovi{\'c}}, {Badman}, {Landi}, {Matteini}, {Bale},
  {Bonnell}, {Case}, {Dudok de Wit}, {Goetz}, {Harvey}, {Kasper}, {Korreck},
  {Livi}, {MacDowall}, {Malaspina}, {Pulupa}, \& {Stevens}}]{2020Bercic}
{Ber{\v{c}}i{\v{c}}}, L., {Larson}, D., {Whittlesey}, P., {et~al.} 2020, \apj,
  892, 88

\bibitem[{{Cane}(1979)}]{1979Cane}
{Cane}, H.~V. 1979, \mnras, 189, 465

\bibitem[{{Case} {et~al.}(2020){Case}, {Kasper}, {Stevens}, {Korreck},
  {Paulson}, {Daigneau}, {Caldwell}, {Freeman}, {Henry}, {Klingensmith},
  {Bookbinder}, {Robinson}, {Berg}, {Tiu}, {Wright}, {Reinhart}, {Curtis},
  {Ludlam}, {Larson}, {Whittlesey}, {Livi}, {Klein}, \&
  {Martinovi{\'c}}}]{Case2020}
{Case}, A.~W., {Kasper}, J.~C., {Stevens}, M.~L., {et~al.} 2020, \apjs, 246, 43

\bibitem[{{Chateau} \& {Meyer-Vernet}(1991)}]{1991Chateau}
{Chateau}, Y.~F. \& {Meyer-Vernet}, N. 1991, \jgr, 96, 5825

\bibitem[{{Chen} {et~al.}(2021){Chen}, {Hu}, {Zhao}, {Kasper}, \&
  {Huang}}]{2021Chen}
{Chen}, Y., {Hu}, Q., {Zhao}, L., {Kasper}, J.~C., \& {Huang}, J. 2021, \apj,
  914, 108

\bibitem[{{D{\'e}moulin}(2009)}]{2009Demoulin}
{D{\'e}moulin}, P. 2009, \solphys, 257, 169

\bibitem[{{Dudok de Wit} {et~al.}(2020){Dudok de Wit}, {Krasnoselskikh},
  {Bale}, {Bonnell}, {Bowen}, {Chen}, {Froment}, {Goetz}, {Harvey},
  {Jagarlamudi}, {Larosa}, {MacDowall}, {Malaspina}, {Matthaeus}, {Pulupa},
  {Velli}, \& {Whittlesey}}]{2020DudokdeWit}
{Dudok de Wit}, T., {Krasnoselskikh}, V.~V., {Bale}, S.~D., {et~al.} 2020,
  \apjs, 246, 39

\bibitem[{{Fargette} {et~al.}(2021){Fargette}, {Lavraud}, {Rouillard},
  {R{\'e}ville}, {Dudok De Wit}, {Froment}, {Halekas}, {Phan}, {Malaspina},
  {Bale}, {Kasper}, {Louarn}, {Case}, {Korreck}, {Larson}, {Pulupa}, {Stevens},
  {Whittlesey}, \& {Berthomier}}]{2021Fargette}
{Fargette}, N., {Lavraud}, B., {Rouillard}, A.~P., {et~al.} 2021, \apj, 919, 96

\bibitem[{{Fox} {et~al.}(2016){Fox}, {Velli}, {Bale}, {Decker}, {Driesman},
  {Howard}, {Kasper}, {Kinnison}, {Kusterer}, {Lario}, {Lockwood}, {McComas},
  {Raouafi}, \& {Szabo}}]{2016Fox}
{Fox}, N.~J., {Velli}, M.~C., {Bale}, S.~D., {et~al.} 2016, \ssr, 204, 7

\bibitem[{{Griton} {et~al.}(2021){Griton}, {Rouillard}, {Poirier}, {Issautier},
  {Moncuquet}, \& {Pinto}}]{2021Griton}
{Griton}, L., {Rouillard}, A.~P., {Poirier}, N., {et~al.} 2021, \apj, 910, 63

\bibitem[{{Halekas} {et~al.}(2022){Halekas}, {Whittlesey}, {Larson},
  {Maksimovic}, {Livi}, {Berthomier}, {Kasper}, {Case}, {Stevens}, {Bale},
  {MacDowall}, \& {Pulupa}}]{2022Halekas}
{Halekas}, J.~S., {Whittlesey}, P., {Larson}, D.~E., {et~al.} 2022, \apj, 936,
  53

\bibitem[{{Halekas} {et~al.}(2020){Halekas}, {Whittlesey}, {Larson},
  {McGinnis}, {Maksimovic}, {Berthomier}, {Kasper}, {Case}, {Korreck},
  {Stevens}, {Klein}, {Bale}, {MacDowall}, {Pulupa}, {Malaspina}, {Goetz}, \&
  {Harvey}}]{2020Halekasa}
{Halekas}, J.~S., {Whittlesey}, P., {Larson}, D.~E., {et~al.} 2020, \apjs, 246,
  22

\bibitem[{{Halekas} {et~al.}(2021){Halekas}, {Whittlesey}, {Larson},
  {McGinnis}, {Bale}, {Berthomier}, {Case}, {Chandran}, {Kasper}, {Klein},
  {Korreck}, {Livi}, {MacDowall}, {Maksimovic}, {Malaspina}, {Matteini},
  {Pulupa}, \& {Stevens}}]{2020Halekasb}
{Halekas}, J.~S., {Whittlesey}, P.~L., {Larson}, D.~E., {et~al.} 2021, \aap,
  650, A15

\bibitem[{{Hellinger} {et~al.}(2013){Hellinger}, {Tr{\'a}Vn{\'\i}{\v{c}}ek},
  {{\v{S}}tver{\'a}k}, {Matteini}, \& {Velli}}]{2013Hellinger}
{Hellinger}, P., {Tr{\'a}Vn{\'\i}{\v{c}}ek}, P.~M., {{\v{S}}tver{\'a}k},
  {\v{S}}., {Matteini}, L., \& {Velli}, M. 2013, Journal of Geophysical
  Research (Space Physics), 118, 1351

\bibitem[{{Hess} {et~al.}(2020){Hess}, {Rouillard}, {Kouloumvakos}, {Liewer},
  {Zhang}, {Dhakal}, {Stenborg}, {Colaninno}, \& {Howard}}]{2020Hess}
{Hess}, P., {Rouillard}, A.~P., {Kouloumvakos}, A., {et~al.} 2020, \apjs, 246,
  25

\bibitem[{{Issautier} {et~al.}(2001{\natexlab{a}}){Issautier}, {Hoang},
  {Moncuquet}, \& {Meyer-Vernet}}]{2001Issautier}
{Issautier}, K., {Hoang}, S., {Moncuquet}, M., \& {Meyer-Vernet}, N.
  2001{\natexlab{a}}, \ssr, 97, 105

\bibitem[{{Issautier} {et~al.}(2008){Issautier}, {Le Chat}, {Meyer-Vernet},
  {Moncuquet}, {Hoang}, {MacDowall}, \& {McComas}}]{2008Issautier}
{Issautier}, K., {Le Chat}, G., {Meyer-Vernet}, N., {et~al.} 2008, \grl, 35,
  L19101

\bibitem[{{Issautier} {et~al.}(1998){Issautier}, {Meyer-Vernet}, {Moncuquet},
  \& {Hoang}}]{1998Issautier}
{Issautier}, K., {Meyer-Vernet}, N., {Moncuquet}, M., \& {Hoang}, S. 1998,
  \jgr, 103, 1969

\bibitem[{{Issautier} {et~al.}(1999{\natexlab{a}}){Issautier}, {Meyer-Vernet},
  {Moncuquet}, \& {Hoang}}]{1999Issautierb}
{Issautier}, K., {Meyer-Vernet}, N., {Moncuquet}, M., \& {Hoang}, S.
  1999{\natexlab{a}}, in American Institute of Physics Conference Series, Vol.
  471, Solar Wind Nine, ed. S.~R. {Habbal}, R.~{Esser}, J.~V. {Hollweg}, \&
  P.~A. {Isenberg}, 581--584

\bibitem[{{Issautier} {et~al.}(1999{\natexlab{b}}){Issautier}, {Meyer-Vernet},
  {Moncuquet}, {Hoang}, \& {McComas}}]{1999Issautier}
{Issautier}, K., {Meyer-Vernet}, N., {Moncuquet}, M., {Hoang}, S., \&
  {McComas}, D.~J. 1999{\natexlab{b}}, \jgr, 104, 6691

\bibitem[{{Issautier} {et~al.}(2001{\natexlab{b}}){Issautier}, {Meyer-Vernet},
  {Pierrard}, \& {Lemaire}}]{2001Issautierb}
{Issautier}, K., {Meyer-Vernet}, N., {Pierrard}, V., \& {Lemaire}, J.
  2001{\natexlab{b}}, \apss, 277, 189

\bibitem[{{Issautier} {et~al.}(2005){Issautier}, {Perche}, {Hoang}, {Lacombe},
  {Maksimovic}, {Bougeret}, \& {Salem}}]{2005Issautier}
{Issautier}, K., {Perche}, C., {Hoang}, S., {et~al.} 2005, Advances in Space
  Research, 35, 2141

\bibitem[{{Issautier} {et~al.}(2001{\natexlab{c}}){Issautier}, {Skoug},
  {Gosling}, {Gary}, \& {McComas}}]{2001Issautierc}
{Issautier}, K., {Skoug}, R.~M., {Gosling}, J.~T., {Gary}, S.~P., \& {McComas},
  D.~J. 2001{\natexlab{c}}, \jgr, 106, 15665

\bibitem[{{Kasper} {et~al.}(2016){Kasper}, {Abiad}, {Austin}, {Balat-Pichelin},
  {Bale}, {Belcher}, {Berg}, {Bergner}, {Berthomier}, {Bookbinder}, {Brodu},
  {Caldwell}, {Case}, {Chand ran}, {Cheimets}, {Cirtain}, {Cranmer}, {Curtis},
  {Daigneau}, {Dalton}, {Dasgupta}, {DeTomaso}, {Diaz-Aguado}, {Djordjevic},
  {Donaskowski}, {Effinger}, {Florinski}, {Fox}, {Freeman}, {Gallagher},
  {Gary}, {Gauron}, {Gates}, {Goldstein}, {Golub}, {Gordon}, {Gurnee}, {Guth},
  {Halekas}, {Hatch}, {Heerikuisen}, {Ho}, {Hu}, {Johnson}, {Jordan},
  {Korreck}, {Larson}, {Lazarus}, {Li}, {Livi}, {Ludlam}, {Maksimovic},
  {McFadden}, {Marchant}, {Maruca}, {McComas}, {Messina}, {Mercer}, {Park},
  {Peddie}, {Pogorelov}, {Reinhart}, {Richardson}, {Robinson}, {Rosen},
  {Skoug}, {Slagle}, {Steinberg}, {Stevens}, {Szabo}, {Taylor}, {Tiu}, {Turin},
  {Velli}, {Webb}, {Whittlesey}, {Wright}, {Wu}, \& {Zank}}]{Kasper2016}
{Kasper}, J.~C., {Abiad}, R., {Austin}, G., {et~al.} 2016, \ssr, 204, 131

\bibitem[{{Kasper} {et~al.}(2019){Kasper}, {Bale}, {Belcher}, {Berthomier},
  {Case}, {Chandran}, {Curtis}, {Gallagher}, {Gary}, {Golub}, {Halekas}, {Ho},
  {Horbury}, {Hu}, {Huang}, {Klein}, {Korreck}, {Larson}, {Livi}, {Maruca},
  {Lavraud}, {Louarn}, {Maksimovic}, {Martinovic}, {McGinnis}, {Pogorelov},
  {Richardson}, {Skoug}, {Steinberg}, {Stevens}, {Szabo}, {Velli},
  {Whittlesey}, {Wright}, {Zank}, {MacDowall}, {McComas}, {McNutt}, {Pulupa},
  {Raouafi}, \& {Schwadron}}]{2019Kasper}
{Kasper}, J.~C., {Bale}, S.~D., {Belcher}, J.~W., {et~al.} 2019, \nat, 576, 228

\bibitem[{{Kasper} {et~al.}(2021){Kasper}, {Klein}, {Lichko}, {Huang}, {Chen},
  {Badman}, {Bonnell}, {Whittlesey}, {Livi}, {Larson}, {Pulupa}, {Rahmati},
  {Stansby}, {Korreck}, {Stevens}, {Case}, {Bale}, {Maksimovic}, {Moncuquet},
  {Goetz}, {Halekas}, {Malaspina}, {Raouafi}, {Szabo}, {MacDowall}, {Velli},
  {Dudok de Wit}, \& {Zank}}]{2021Kasper}
{Kasper}, J.~C., {Klein}, K.~G., {Lichko}, E., {et~al.} 2021, \prl, 127, 255101

\bibitem[{{Korreck} {et~al.}(2020){Korreck}, {Szabo}, {Nieves Chinchilla},
  {Lavraud}, {Luhmann}, {Niembro}, {Higginson}, {Alzate}, {Wallace}, {Paulson},
  {Rouillard}, {Kouloumvakos}, {Poirier}, {Kasper}, {Case}, {Stevens}, {Bale},
  {Pulupa}, {Whittlesey}, {Livi}, {Goetz}, {Larson}, {Malaspina}, {Morgan},
  {Narock}, {Schwadron}, {Bonnell}, {Harvey}, \& {Wygant}}]{2020Korreck}
{Korreck}, K.~E., {Szabo}, A., {Nieves Chinchilla}, T., {et~al.} 2020, \apjs,
  246, 69

\bibitem[{{Le Chat} {et~al.}(2011){Le Chat}, {Issautier}, {Meyer-Vernet}, \&
  {Hoang}}]{2011LeChat}
{Le Chat}, G., {Issautier}, K., {Meyer-Vernet}, N., \& {Hoang}, S. 2011,
  \solphys, 271, 141

\bibitem[{{Liu} {et~al.}(2021{\natexlab{a}}){Liu}, {Issautier}, {Meyer-Vernet},
  {Moncuquet}, {Maksimovic}, {Halekas}, {Huang}, {Griton}, {Bale}, {Bonnell},
  {Case}, {Goetz}, {Harvey}, {Kasper}, {MacDowall}, {Malaspina}, {Pulupa}, \&
  {Stevens}}]{2021Liu}
{Liu}, M., {Issautier}, K., {Meyer-Vernet}, N., {et~al.} 2021{\natexlab{a}},
  \aap, 650, A14

\bibitem[{{Liu} {et~al.}(2020){Liu}, {Issautier}, {Meyer-Vernet}, {Moncuquet},
  {Maksimovic}, {Kasper}, {Halekas}, {Griton}, {Huang}, {Bale}, \&
  {Pulupa}}]{2020Liu}
{Liu}, M., {Issautier}, K., {Meyer-Vernet}, N., {et~al.} 2020, in AGU Fall
  Meeting Abstracts, Vol. 2020, SH052--04

\bibitem[{{Liu} {et~al.}(2021{\natexlab{b}}){Liu}, {Chen}, {Stevens}, \&
  {Liu}}]{2021ApJLiu}
{Liu}, Y.~D., {Chen}, C., {Stevens}, M.~L., \& {Liu}, M. 2021{\natexlab{b}},
  \apjl, 908, L41

\bibitem[{{Livi} {et~al.}(2022){Livi}, {Larson}, {Kasper}, {Abiad}, {Case},
  {Klein}, {Curtis}, {Dalton}, {Stevens}, {Korreck}, {Ho}, {Robinson}, {Tiu},
  {Whittlesey}, {Verniero}, {Halekas}, {McFadden}, {Marckwordt}, {Slagle},
  {Abatcha}, {Rahmati}, \& {McManus}}]{2022Livi}
{Livi}, R., {Larson}, D.~E., {Kasper}, J.~C., {et~al.} 2022, \apj, 938, 138

\bibitem[{{Lopez} \& {Freeman}(1986)}]{1986Lopez}
{Lopez}, R.~E. \& {Freeman}, J.~W. 1986, \jgr, 91, 1701

\bibitem[{{Lund} {et~al.}(1994){Lund}, {Labelle}, \& {Treumann}}]{1994Lund}
{Lund}, E.~J., {Labelle}, J., \& {Treumann}, R.~A. 1994, \jgr, 99, 23651

\bibitem[{{Maksimovic} {et~al.}(2020){Maksimovic}, {Bale},
  {Ber{\v{c}}i{\v{c}}}, {Bonnell}, {Case}, {Wit}, {Goetz}, {Halekas}, {Harvey},
  {Issautier}, {Kasper}, {Korreck}, {Jagarlamudi}, {Lahmiti}, {Larson},
  {Lecacheux}, {Livi}, {MacDowall}, {Malaspina}, {Martinovi{\'c}},
  {Meyer-Vernet}, {Moncuquet}, {Pulupa}, {Salem}, {Stevens},
  {{\v{S}}tver{\'a}k}, {Velli}, \& {Whittlesey}}]{Maksimovic2020}
{Maksimovic}, M., {Bale}, S.~D., {Ber{\v{c}}i{\v{c}}}, L., {et~al.} 2020,
  \apjs, 246, 62

\bibitem[{{Maksimovic} {et~al.}(2000){Maksimovic}, {Gary}, \&
  {Skoug}}]{2000Maksimovic}
{Maksimovic}, M., {Gary}, S.~P., \& {Skoug}, R.~M. 2000, \jgr, 105, 18337

\bibitem[{{Maksimovic} {et~al.}(1995){Maksimovic}, {Hoang}, {Meyer-Vernet},
  {Moncuquet}, {Bougeret}, {Phillips}, \& {Canu}}]{1995Maksimovic}
{Maksimovic}, M., {Hoang}, S., {Meyer-Vernet}, N., {et~al.} 1995, \jgr, 100,
  19881

\bibitem[{{Maksimovic} {et~al.}(2005{\natexlab{a}}){Maksimovic}, {Issautier},
  {Meyer-Vernet}, {Perche}, {Moncuquet}, {Zouganelis}, {Bale}, {Vilmer}, \&
  {Bougeret}}]{2005Maksimovic_AdSpR}
{Maksimovic}, M., {Issautier}, K., {Meyer-Vernet}, N., {et~al.}
  2005{\natexlab{a}}, Advances in Space Research, 36, 1471

\bibitem[{{Maksimovic} {et~al.}(2001){Maksimovic}, {Pierrard}, \&
  {Lemaire}}]{2001Maksimovic}
{Maksimovic}, M., {Pierrard}, V., \& {Lemaire}, J. 2001, \apss, 277, 181

\bibitem[{{Maksimovic} {et~al.}(1997){Maksimovic}, {Pierrard}, \&
  {Lemaire}}]{1997Maksimovic}
{Maksimovic}, M., {Pierrard}, V., \& {Lemaire}, J.~F. 1997, \aap, 324, 725

\bibitem[{Maksimovic {et~al.}(2021)Maksimovic, Walsh, Pierrard,
  {\v{S}}tver{\'a}k, \& Zouganelis}]{Maksimovic2021}
Maksimovic, M., Walsh, A.~P., Pierrard, V., {\v{S}}tver{\'a}k, {\v{S}}., \&
  Zouganelis, I. 2021, Electron Kappa Distributions in the Solar Wind: Cause of
  the Acceleration or Consequence of the Expansion?, ed. M.~Lazar \&
  H.~Fichtner (Cham: Springer International Publishing), 39--51

\bibitem[{{Maksimovic} {et~al.}(2005{\natexlab{b}}){Maksimovic}, {Zouganelis},
  {Chaufray}, {Issautier}, {Scime}, {Littleton}, {Marsch}, {McComas}, {Salem},
  {Lin}, \& {Elliott}}]{2005Maksimovic}
{Maksimovic}, M., {Zouganelis}, I., {Chaufray}, J.~Y., {et~al.}
  2005{\natexlab{b}}, Journal of Geophysical Research (Space Physics), 110,
  A09104

\bibitem[{{Manning} \& {Dulk}(2001)}]{2001Manning}
{Manning}, R. \& {Dulk}, G.~A. 2001, \aap, 372, 663

\bibitem[{{Markwardt}(2009)}]{2009Markwardt}
{Markwardt}, C.~B. 2009, in Astronomical Society of the Pacific Conference
  Series, Vol. 411, Astronomical Data Analysis Software and Systems XVIII, ed.
  D.~A. {Bohlender}, D.~{Durand}, \& P.~{Dowler}, 251

\bibitem[{{Marsch} {et~al.}(1989){Marsch}, {Pilipp}, {Thieme}, \&
  {Rosenbauer}}]{1989Marsch}
{Marsch}, E., {Pilipp}, W.~G., {Thieme}, K.~M., \& {Rosenbauer}, H. 1989, \jgr,
  94, 6893

\bibitem[{{Martinovi{\'c}} {et~al.}(2022){Martinovi{\'c}}, {Dordevi{\'c}},
  {Klein}, {Maksimovi{\'c}}, {Issautier}, {Liu}, {Pulupa}, {Bale}, {Halekas},
  \& {McManus}}]{2022Martinovic}
{Martinovi{\'c}}, M.~M., {Dordevi{\'c}}, A.~R., {Klein}, K.~G., {et~al.} 2022,
  Journal of Geophysical Research (Space Physics), 127, e30182

\bibitem[{{Martinovi{\'c}} {et~al.}(2020){Martinovi{\'c}}, {Klein}, {Gramze},
  {Jain}, {Maksimovi{\'c}}, {Zaslavsky}, {Salem}, {Zouganelis}, \&
  {Simi{\'c}}}]{2020Martinovic}
{Martinovi{\'c}}, M.~M., {Klein}, K.~G., {Gramze}, S.~R., {et~al.} 2020,
  Journal of Geophysical Research (Space Physics), 125, e28113

\bibitem[{{Martinovi{\'c}} {et~al.}(2021){Martinovi{\'c}}, {Klein}, {Huang},
  {Chandran}, {Kasper}, {Lichko}, {Bowen}, {Chen}, {Matteini}, {Stevens},
  {Case}, \& {Bale}}]{2021Martinovic}
{Martinovi{\'c}}, M.~M., {Klein}, K.~G., {Huang}, J., {et~al.} 2021, \apj, 912,
  28

\bibitem[{{Martinovi{\'c}} {et~al.}(2016){Martinovi{\'c}}, {Zaslavsky},
  {Maksimovi{\'c}}, {Meyer-Vernet}, {{\r{A}} egan}, {Zouganelis}, {Salem},
  {Pulupa}, \& {Bale}}]{2016Martinovic}
{Martinovi{\'c}}, M.~M., {Zaslavsky}, A., {Maksimovi{\'c}}, M., {et~al.} 2016,
  Journal of Geophysical Research (Space Physics), 121, 129

\bibitem[{{Matthaeus} {et~al.}(2006){Matthaeus}, {Elliott}, \&
  {McComas}}]{2006Matthaeus}
{Matthaeus}, W.~H., {Elliott}, H.~A., \& {McComas}, D.~J. 2006, Journal of
  Geophysical Research (Space Physics), 111, A10103

\bibitem[{{Meyer-Vernet}(1979)}]{1979meyervernet}
{Meyer-Vernet}, N. 1979, \jgr, 84, 5373

\bibitem[{{Meyer-Vernet} {et~al.}(1986){Meyer-Vernet}, {Couturier}, {Hoang},
  {Perche}, {Steinberg}, {Fainberg}, \& {Meetre}}]{1986meyervernet}
{Meyer-Vernet}, N., {Couturier}, P., {Hoang}, S., {et~al.} 1986, Science, 232,
  370

\bibitem[{{Meyer-Vernet} {et~al.}(1993){Meyer-Vernet}, {Hoang}, \&
  {Moncuquet}}]{1993MeyerVernet}
{Meyer-Vernet}, N., {Hoang}, S., \& {Moncuquet}, M. 1993, \jgr, 98, 21163

\bibitem[{{Meyer-Vernet} \& {Issautier}(1998)}]{1998meyervernet}
{Meyer-Vernet}, N. \& {Issautier}, K. 1998, \jgr, 103, 29705

\bibitem[{Meyer-Vernet {et~al.}(2017)Meyer-Vernet, Issautier, \&
  Moncuquet}]{2017meyervernet}
Meyer-Vernet, N., Issautier, K., \& Moncuquet, M. 2017, \jgr, 122, 7925

\bibitem[{{Meyer-Vernet} {et~al.}(2022){Meyer-Vernet}, {Lecacheux},
  {Issautier}, \& {Moncuquet}}]{2022meyervernet}
{Meyer-Vernet}, N., {Lecacheux}, A., {Issautier}, K., \& {Moncuquet}, M. 2022,
  \aap, 658, L12

\bibitem[{{Meyer-Vernet} {et~al.}(2003){Meyer-Vernet}, {Mangeney},
  {Maksimovic}, {Pantellini}, \& {Issautier}}]{2003meyervernet}
{Meyer-Vernet}, N., {Mangeney}, A., {Maksimovic}, M., {Pantellini}, F., \&
  {Issautier}, K. 2003, in American Institute of Physics Conference Series,
  Vol. 679, Solar Wind Ten, ed. M.~{Velli}, R.~{Bruno}, F.~{Malara}, \&
  B.~{Bucci}, 263--266

\bibitem[{{Meyer-Vernet} \& {Moncuquet}(2020)}]{2020meyervernet}
{Meyer-Vernet}, N. \& {Moncuquet}, M. 2020, Journal of Geophysical Research
  (Space Physics), 125, e27723

\bibitem[{{Meyer-Vernet} \& {Perche}(1989)}]{1989meyervernet}
{Meyer-Vernet}, N. \& {Perche}, C. 1989, \jgr, 94, 2405

\bibitem[{{Moncuquet} {et~al.}(2005){Moncuquet}, {Lecacheux}, {Meyer-Vernet},
  {Cecconi}, \& {Kurth}}]{2005Moncuquet}
{Moncuquet}, M., {Lecacheux}, A., {Meyer-Vernet}, N., {Cecconi}, B., \&
  {Kurth}, W.~S. 2005, \grl, 32, L20S02

\bibitem[{{Moncuquet} {et~al.}(2006){Moncuquet}, {Matsumoto}, {Bougeret},
  {Blomberg}, {Issautier}, {Kasaba}, {Kojima}, {Maksimovic}, {Meyer-Vernet}, \&
  {Zarka}}]{2006Moncuquet}
{Moncuquet}, M., {Matsumoto}, H., {Bougeret}, J.~L., {et~al.} 2006, Advances in
  Space Research, 38, 680

\bibitem[{{Moncuquet} {et~al.}(1995){Moncuquet}, {Meyer-Vernet}, \&
  {Hoang}}]{1995Moncuquet}
{Moncuquet}, M., {Meyer-Vernet}, N., \& {Hoang}, S. 1995, \jgr, 100, 21697

\bibitem[{{Moncuquet} {et~al.}(1997){Moncuquet}, {Meyer-Vernet}, {Hoang},
  {Forsyth}, \& {Canu}}]{1997Moncuquet}
{Moncuquet}, M., {Meyer-Vernet}, N., {Hoang}, S., {Forsyth}, R.~J., \& {Canu},
  P. 1997, \jgr, 102, 2373

\bibitem[{{Moncuquet} {et~al.}(2020){Moncuquet}, {Meyer-Vernet}, {Issautier},
  {Pulupa}, {Bonnell}, {Bale}, {Dudok de Wit}, {Goetz}, {Griton}, {Harvey},
  {MacDowall}, {Maksimovic}, \& {Malaspina}}]{2020Moncuquet}
{Moncuquet}, M., {Meyer-Vernet}, N., {Issautier}, K., {et~al.} 2020, \apjs,
  246, 44

\bibitem[{{Novaco} \& {Brown}(1978)}]{1978Novaco}
{Novaco}, J.~C. \& {Brown}, L.~W. 1978, \apj, 221, 114

\bibitem[{{Pierrard} {et~al.}(2020){Pierrard}, {Lazar}, \&
  {{\v{S}}tver{\'a}k}}]{2020Pierrard}
{Pierrard}, V., {Lazar}, M., \& {{\v{S}}tver{\'a}k}, S. 2020, \solphys, 295,
  151

\bibitem[{{Pilipp} {et~al.}(1990){Pilipp}, {Muehlhaeuser}, {Miggenrieder},
  {Rosenbauer}, \& {Schwenn}}]{1990Pilipp}
{Pilipp}, W.~G., {Muehlhaeuser}, K.~H., {Miggenrieder}, H., {Rosenbauer}, H.,
  \& {Schwenn}, R. 1990, \jgr, 95, 6305

\bibitem[{{Pulupa} {et~al.}(2020){Pulupa}, {Bale}, {Badman}, {Bonnell}, {Case},
  {de Wit}, {Goetz}, {Harvey}, {Hegedus}, {Kasper}, {Korreck},
  {Krasnoselskikh}, {Larson}, {Lecacheux}, {Livi}, {MacDowall}, {Maksimovic},
  {Malaspina}, {Mart{\'\i}nez Oliveros}, {Meyer-Vernet}, {Moncuquet},
  {Stevens}, \& {Whittlesey}}]{2020Pulupa}
{Pulupa}, M., {Bale}, S.~D., {Badman}, S.~T., {et~al.} 2020, \apjs, 246, 49

\bibitem[{{Pulupa} {et~al.}(2017){Pulupa}, {Bale}, {Bonnell}, {Bowen},
  {Carruth}, {Goetz}, {Gordon}, {Harvey}, {Maksimovic},
  {Mart{\'\i}nez-Oliveros}, {Moncuquet}, {Saint-Hilaire}, {Seitz}, \&
  {Sundkvist}}]{2017Pulupa}
{Pulupa}, M., {Bale}, S.~D., {Bonnell}, J.~W., {et~al.} 2017, \jgr, 122, 2836

\bibitem[{{Salem} {et~al.}(2001){Salem}, {Bosqued}, {Larson}, {Mangeney},
  {Maksimovic}, {Perche}, {Lin}, \& {Bougeret}}]{2001salem}
{Salem}, C., {Bosqued}, J.~M., {Larson}, D.~E., {et~al.} 2001, \jgr, 106, 21701

\bibitem[{{Schippers} {et~al.}(2013){Schippers}, {Moncuquet}, {Meyer-Vernet},
  \& {Lecacheux}}]{2013Schippers}
{Schippers}, P., {Moncuquet}, M., {Meyer-Vernet}, N., \& {Lecacheux}, A. 2013,
  Journal of Geophysical Research (Space Physics), 118, 7170

\bibitem[{{Totten} {et~al.}(1995){Totten}, {Freeman}, \& {Arya}}]{1995Totten}
{Totten}, T.~L., {Freeman}, J.~W., \& {Arya}, S. 1995, \jgr, 100, 13

\bibitem[{{{\v{S}}tver{\'a}k} {et~al.}(2009){{\v{S}}tver{\'a}k}, {Maksimovic},
  {Tr{\'a}vn{\'\i}{\v{c}}ek}, {Marsch}, {Fazakerley}, \& {Scime}}]{2009JGRA}
{{\v{S}}tver{\'a}k}, {\v{S}}., {Maksimovic}, M., {Tr{\'a}vn{\'\i}{\v{c}}ek},
  P.~M., {et~al.} 2009, \jgr, 114, A05104

\bibitem[{{{\v{S}}tver{\'a}k} {et~al.}(2015){{\v{S}}tver{\'a}k},
  {Tr{\'a}vn{\'\i}{\v{c}}ek}, \& {Hellinger}}]{2015JGRA}
{{\v{S}}tver{\'a}k}, {\v{S}}., {Tr{\'a}vn{\'\i}{\v{c}}ek}, P.~M., \&
  {Hellinger}, P. 2015, Journal of Geophysical Research (Space Physics), 120,
  8177

\bibitem[{{Wang}(2010)}]{2010Wang}
{Wang}, Y.~M. 2010, \apjl, 715, L121

\bibitem[{{Wang} \& {Sheeley}(1990)}]{1990Wang}
{Wang}, Y.~M. \& {Sheeley}, N.~R., J. 1990, \apj, 355, 726

\bibitem[{Whittlesey {et~al.}(2020)Whittlesey, Larson, Kasper, Halekas,
  Abatcha, Abiad, Berthomier, Case, Chen, Curtis, Dalton, Klein, Korreck, Livi,
  Ludlam, Marckwordt, Rahmati, Robinson, Slagle, Stevens, Tiu, \&
  Verniero}]{Whittlesey_2020}
Whittlesey, P.~L., Larson, D.~E., Kasper, J.~C., {et~al.} 2020, \apjs, 246, 74

\bibitem[{{Wilson} {et~al.}(2018){Wilson}, {Stevens}, {Kasper}, {Klein},
  {Maruca}, {Bale}, {Bowen}, {Pulupa}, \& {Salem}}]{2018Wilson}
{Wilson}, Lynn~B., I., {Stevens}, M.~L., {Kasper}, J.~C., {et~al.} 2018, \apjs,
  236, 41

\bibitem[{{Woodham} {et~al.}(2021){Woodham}, {Horbury}, {Matteini}, {Woolley},
  {Laker}, {Bale}, {Nicolaou}, {Stawarz}, {Stansby}, {Hietala}, {Larson},
  {Livi}, {Verniero}, {McManus}, {Kasper}, {Korreck}, {Raouafi}, {Moncuquet},
  \& {Pulupa}}]{2021Woodham}
{Woodham}, L.~D., {Horbury}, T.~S., {Matteini}, L., {et~al.} 2021, \aap, 650,
  L1

\bibitem[{{Zaslavsky} {et~al.}(2011){Zaslavsky}, {Meyer-Vernet}, {Hoang},
  {Maksimovic}, \& {Bale}}]{2011Zaslavsky}
{Zaslavsky}, A., {Meyer-Vernet}, N., {Hoang}, S., {Maksimovic}, M., \& {Bale},
  S.~D. 2011, Radio Science, 46, RS2008

\bibitem[{{Zhao} {et~al.}(2020){Zhao}, {Zank}, {Adhikari}, {Hu}, {Kasper},
  {Bale}, {Korreck}, {Case}, {Stevens}, {Bonnell}, {Dudok de Wit}, {Goetz},
  {Harvey}, {MacDowall}, {Malaspina}, {Pulupa}, {Larson}, {Livi}, {Whittlesey},
  \& {Klein}}]{2020Zhao}
{Zhao}, L.~L., {Zank}, G.~P., {Adhikari}, L., {et~al.} 2020, \apjs, 246, 26

\bibitem[{{Zhao} {et~al.}(2021){Zhao}, {Yan}, {Liu}, {Liu}, \&
  {Shi}}]{2021Zhao}
{Zhao}, S.~Q., {Yan}, H., {Liu}, T.~Z., {Liu}, M., \& {Shi}, M. 2021, \apj,
  923, 253

\bibitem[{{Zhou} {et~al.}(2022){Zhou}, {Xu}, {Zuo}, {Wang}, {Wang}, {Ye},
  {Wang}, {Chang}, {Wang}, \& {Luo}}]{2022Zhou}
{Zhou}, Z., {Xu}, X., {Zuo}, P., {et~al.} 2022, \grl, 49, e97564

\bibitem[{{Zouganelis} {et~al.}(2004){Zouganelis}, {Maksimovic},
  {Meyer-Vernet}, {Lamy}, \& {Issautier}}]{2004Zouganelis}
{Zouganelis}, I., {Maksimovic}, M., {Meyer-Vernet}, N., {Lamy}, H., \&
  {Issautier}, K. 2004, \apj, 606, 542

\end{thebibliography}
\bibliographystyle{aa} 

\end{document}